\documentclass[usenatbib,usegraphicx,useAMS]{mn2e}
\pdfoutput=1
\usepackage{graphicx}
\usepackage{subfig}
\usepackage{verbatim}
\usepackage[usenames,dvipsnames]{color}
\usepackage{booktabs}
\usepackage{dsfont}
\usepackage{listings} 
\usepackage{amsmath}
\usepackage{soul}
\usepackage{amssymb}
\usepackage{pifont}
\usepackage[point]{rccol}

\allowdisplaybreaks[1]

\newcommand{\App}[1]{Appendix~\ref{#1}}

\newcommand{\Eq}[1]{Eq.~\ref{#1}}
\newcommand{\Eeq}[1]{Eq.~(\ref{#1})}

\newcommand{\Fig}[1]{Fig.~\ref{#1}}
\newcommand{\Sec}[1]{\S\ref{#1}}
\newcommand{\Tab}[1]{Table~\ref{#1}}

\newcommand{\cmark}{\ding{51}}%

\def\ramses    {{\sc ramses}}

\def\ramsesrt  {{\sc ramses-rt}}

\def\dw {{\sc g8}}
\def\sbc {{\sc g9}}
\def\mw {{\sc g10}}
\def\sima {{\sc nofb}}
\def\simb {{\sc sn}}
\def\simc {{\sc sn\_rh}}
\def\simd {{\sc sn\_rhp}}
\def\sime {{\sc sn\_rhpd}}
\def\simf {{\sc rhpd}}

\def\hei    {{\rm{He\textsc{i}}}}
\def\heii   {\rm{He\textsc{ii}}}

\def\hi    {{\rm{H\textsc{i}}}}
\def\hii   {\rm{H\textsc{ii}}}

\def\nh {n_{\rm{H}}}

\def\cci {{\rm{cm}}^{-3}}                                   
\def\cc {\rm{cm}^{3}}                                       
\def\ccg {{\rm{cm}}^{2} \, \rm{g}^{-1}}                     
\def\cs {\rm{cm}^{2}}                                       
\def\ergs {\rm{erg} \, \rm{s}^{-1}}                         
\def\kms {\rm{km} \, \rm{s}^{-1}}                           
\def\msy {\Msun/\rm{yr}}                                    
\def\pc {\rm{pc}}                                           
\def\sm {\rm{s}^{-1}}                                       
\def\csavg {\sigma}           
\def\cred {\tilde c}      
\def\csound {c_{\rm{s}}}

\def\dx {\Delta x}
\def\Dxmax {\Delta x_{\rm max}}
\def\Dxmin {\Delta x_{\rm min}}
\def\eavg {\bar{\epsilon}}   

\def\fgas {f_{\rm gas}}

\def\fred {f_{\gamma}}
\def\kb {k_{\rm B}}
\def\kIR {\kappa_{\rm IR}}
\def\kOIR {\tilde{\kappa}_{{\rm IR}}}
\def\kOi {\tilde{\kappa}_i}
\def\kO {\tilde{\kappa}}
\def\ki {\kappa_i}
\def\kOpt {\kappa_{\rm Opt}}
\def\kOOpt {\tilde{\kappa}_{{\rm Opt}}}
\def\kOUV {\tilde{\kappa}_{{\rm UV}}}
\def\kUV {\kappa_{\rm UV}}
\def\Lbox {L_{\rm box}}
\def\LJeans {\lambda_{\rm J}}
\def\LumNum {\hat{L}}
\def\LumNumSpec {\hat{\mathcal L}}
\def\LumSpec {\mathcal L}

\def\Mbulge {M_{\rm bulge}}
\def\Mdisk {M_{\rm disk}}
\def\mcell {m_{\rm cell}}
\def\Mhalo {M_{\rm halo}}
\def\mp {m_{\rm p}}
\def\Mstar {M_{*}}
\def\mstar {m_{*}}
\def\mstarm {m_{*}^{\rm{max}}}
\def\Msun {\rm M_{\odot}}

\def\Npart {N_{\rm part}}

\def\ns {n_{*}}

\def\Rvir {R_{\rm vir}}
\def\rS {r_{\rm{S}}}        
\def\sfeff {\epsilon_{\rm{ff}}}
\def\SFRdx {{\rm SFR}_{\Delta x}}
\def\stromgren {Str\"omgren}

\def\tauIR {{\tau_{\rm IR}}}

\def\TJeans {T_{\rm{J}}}      
\def\Tnt {T_{\rm{Eff}}}

\def\vcirc {v_{\rm circ}}
\def\Zsun {\rm Z_{\odot}}       
\def\RM {R14}  
\def\RRT {R13}  

\mathchardef\mhyphen="2D

\long\def\symbolfootnote[#1]#2{\begingroup%
\def\thefootnote{\fnsymbol{footnote}}\footnote[#1]{#2}\endgroup}

\begin{document}

\title[Galaxies that Shine]{Galaxies that Shine:
  radiation-hydrodynamical simulations of disk galaxies}
\author[Rosdahl, Schaye, Teyssier \& Agertz]
{Joakim Rosdahl$^{1}$\thanks{E-mail: joki@strw.leidenuniv.nl},
  Joop Schaye$^{1}$, Romain Teyssier$^{2}$, and Oscar Agertz$^{3}$\\
  $^1$Leiden Observatory, Leiden University, P.O. Box 9513, 2300 RA,
  Leiden, The Netherlands \\
  $^2$Institute for Computational Science, University of Z\"urich,
  Winterthurerstrasse 190, CH-8057 Z\"urich, Switzerland \\
  $^3$Department of Physics, University of Surrey, Guildford, GU2 7XH
  Surrey, UK}

\maketitle
\begin{abstract}
  Radiation feedback is typically implemented using subgrid recipes in
  hydrodynamical simulations of galaxies. Very little work has so far
  been performed using radiation-hydrodynamics (RHD), and there is no
  consensus on the importance of radiation feedback in galaxy
  evolution.  We present RHD simulations of isolated galaxy disks of
  different masses with a resolution of 18 pc. Besides accounting for
  supernova feedback, our simulations are the first galaxy-scale
  simulations to include RHD treatments of photo-ionisation heating
  and radiation pressure, from both direct optical/UV radiation and
  multi-scattered, re-processed infrared (IR) radiation. Photo-heating
  smooths and thickens the disks and suppresses star formation about
  as much as the inclusion of (``thermal dump'') supernova feedback
  does.  These effects decrease with galaxy mass and are mainly due to
  the prevention of the formation of dense clouds, as opposed to their
  destruction. Radiation pressure, whether from direct or IR
  radiation, has little effect, but for the IR radiation we show that
  its impact is limited by our inability to resolve the high optical
  depths for which multi-scattering becomes important. While
  artificially boosting the IR optical depths does reduce the star
  formation, it does so by smoothing the gas rather than by generating
  stronger outflows. We conclude that although higher-resolution
  simulations, and potentially also different supernova
  implementations, are needed for confirmation, our findings suggest
  that radiation feedback is more gentle and less effective than is
  often assumed in subgrid prescriptions.
\end{abstract}
\begin{keywords}
  galaxies: evolution -- methods: numerical -- radiative transfer
\end{keywords}

\section{Introduction} \label{Intro.sec} 

To first order, gravity describes the formation of structure in the
Universe \citep{Peebles:1970if, Zeldovich:1970wx}. The formation of
galaxies in dark matter (DM) halos also requires radiative cooling to
relieve pressure and dissipate angular momentum \citep{Binney:1977fp,
  Rees:1977ur, Silk:1977ii}. It is also well established that in order
to halt the collapse of gas into galaxies, dense substructures, and
eventually stars, counteracting feedback processes are required
\citep[e.g.][]{White:1978uk}. Without feedback, galaxies collapse and
form stars too efficiently, compared to observations.

Early simulations focused on feedback in the form of supernovae
\citep[SNe; e.g.][]{Katz:1992gu, Navarro:1993tq} and later active
galactic nuclei \cite[e.g.][]{DiMatteo:2005hl, Booth:2009jt,
  Dubois:2010jv}, where the latter is thought to be dominant in
massive (``$L>L_*$'') galaxies \citep{Bower:2006fj}. However,
simulations that include those feedback processes still struggle to
produce galaxies that match observations in terms of their star
formation histories and morphology, \citep{Scannapieco:2012iy}.

Analytical work by e.g. \cite{Thompson:2005dq} and
\cite{Murray:2005jt, Murray:2010gh, Murray:2011en} suggests that
radiation feedback may be an important missing ingredient. Recent
hydrodynamical simulations therefore often enlist stellar radiation in
their subgrid feedback models \citep[e.g.][]{Brook:2012ht,
  Agertz:2013il, Stinson:2013ex, Roskar:2014be, Ceverino:2014dw,
  Kannan:2014jb, Kannan:2014di, Hopkins:2014dn, Agertz:2015gb}. The
added radiation feedback usually contributes directly to direct
suppression of star formation, and increases galactic outflows, which
can expel the gas altogether and enrich the intergalactic medium (IGM)
with metals.  The idea of radiation feedback has proven so successful
that most cosmological simulations nowadays invoke it in some form,
although the implementations vary a lot, and they are often motivated
empirically rather than physically. Radiation feedback on galactic
scales is usually modelled with subgrid recipes in otherwise purely
hydrodynamical (HD) codes. These HD recipes must make a number of
assumptions about e.g. the absorption of photons, mean free paths, and
shielding. They can thus only to a limited degree be used to
investigate how important radiation is for the formation and evolution
of galaxies, and how the radiation interacts with the baryons,
i.e. how radiation feedback actually works.

The recent literature on simulations of galaxy evolution usually
considers three radiation feedback processes: photoionisation heating
of gas, direct pressure from ionising photons, and indirect pressure
from reprocessed, multi-scattering, infrared (IR) photons.
Simulations often contain only a subset of these processes, and there
is no general consensus on the importance of radiation feedback as a
whole, or on which of these processes dominate under which
circumstances (see \Sec{comparison.sec}).

A more assumption-free and physically correct description of radiation
feedback requires the use of radiation-hydrodynamics (RHD), which
models the emission and propagation of photons and their interaction
with the gas self-consistently. RHD can help tell us if and how
radiation feedback works, and this information can then be used to
improve HD subgrid recipes of radiation feedback.

However, RHD is both complex and costly compared to HD. For the most
part, it has therefore not been used directly in simulations of
structure formation, or more generally in studies of galaxy
evolution. In recent years however, the use of RHD has been on the
rise in computational astronomy, and RHD implementations have evolved
towards being usable in cosmological and galaxy-scale simulations that
resolve the interstellar medium (ISM) \citep{Wise:2012ej, Wise:2012dh,
  Pawlik:2013jz, Wise:2014kt}.

In \citet[][hereafter \RRT{}]{Rosdahl:2013cea} we presented an RHD
implementation in the cosmological code \ramses{}
\citep{Teyssier:2002fj}, which we called \ramsesrt{}. In that paper we
modelled the emission and propagation of photons and their interaction
with hydrogen and helium via ionisation and heating. In
\citet[][hereafter \RM{}]{Rosdahl:2015dj}, we added two aforementioned
processes to the implementation, which are though to be relevant for
galactic feedback: radiation pressure, i.e. momentum transfer from
photons to gas, and the diffusion and trapping of multi-scattered IR
radiation in optically thick gas.

In the present paper, we use the RHD implementation that we have
detailed in the two previous papers to study the effect of stellar
radiation feedback on galactic scales. We use a set of \ramsesrt{}
simulations of isolated galactic disk simulations, where we include
stellar radiation feedback, combined with ``thermal dump'' SN
feedback. The main questions we attempt to answer are:
\begin{itemize}
\item What role does stellar radiation feedback play in regulating
  galaxy evolution, and how does this role vary with the mass and
  metallicity of the galaxy?
\item How does the interplay of radiation and SN feedback work?
  Specifically, does radiation boost the effect of SNe?
\item Where stellar radiation feedback plays a role, what is the
  dominant physical process: photoionisation heating, direct
  pressure from the ionising photons on the gas, or indirect pressure
  via dust particles UV and reprocessed IR radiation?
\end{itemize}

In this paper, we study the effects of turning on the stellar
radiation in galaxies, while making minimal assumptions about what
happens on unresolved scales. While using RHD implies radiation
feedback is modelled from ``first principles'', we stress that it is
still necessary to make a number of approximations, both in the
modelling of the radiation itself and in its interaction with gas and
dust.  Also, and importantly, although we resolve the ISM to some
extent, we do not resolve molecular clouds, the scales at which the
radiation feedback originates, and at which the radiation couples most
efficiently with the gas. We expect the current simulations to give us
hints as to what radiation feedback does in reality, and, equally
importantly, to teach us what improvements in modelling and resolution
are required in future work.

The structure of the paper is as follows. In \Sec{methods.sec} we
present an overview of the code, the setup of galaxy disks of three
masses, and details of the modelling of gas, stellar populations, and
feedback. In \Sec{results.sec} we present the results, where we
successively incorporate SN and radiation feedback processes and
compare their effects on the galaxies. We focus on the suppression of
star formation and the generation of outflows, study \emph{how}
radiation feedback plays a role, and examine trends with galaxy mass
and metallicity. In \Sec{Discussions.sec}, we discuss and justify our
main findings on analytic grounds, demonstrate how they are limited by
resolution, probe what effects we can expect when the resolution is
increased beyond the current limits, and qualitatively compare our
results to previous publications. Finally, in \Sec{Conclusions.sec} we
summarise our main conclusions and discuss interesting future
directions. The appendices provide details on the model we use for
stellar population specific luminosities and convergence tests.

\section{Simulations} \label{methods.sec}
\begin{table*}
  \centering
  \caption
  {Simulation parameters for the three disk galaxies. The listed
    parameters are, from left to right:
    Galaxy acronym used throughout the paper, $\vcirc$: NFW circular
    velocity, for the IC generation, $\Rvir$: halo virial radius
    (defined as the radius at which the DM density is $200$ times the
    critical density at redshift zero), $\Lbox$:
    simulation box length, $\Mhalo$: DM halo mass, $\Mdisk$: disk
    galaxy mass in baryons (stars+gas), $\fgas$: disk gas fraction in the
    ICs, $\Mbulge$: stellar bulge mass in the ICs, $\Npart$: Number of
    DM/stellar particles in the ICs, $\mstar$: mass of stellar particles
    formed during the simulations, $\Dxmax$: coarsest cell resolution,
    $\Dxmin$: finest cell resolution, $Z_{\rm disk}$: disk
    metallicity.}
  \label{sims.tbl}
  \begin{tabular}{l|rrrrrrrrrrrr}
    \toprule
    Galaxy  & $\vcirc$ & $\Rvir$ & $\Lbox$   & $\Mhalo$ 
            & $\Mdisk$ & $\fgas$ & $\Mbulge$ & $\Npart$   
            & $\mstar$ & $\Dxmax$& $\Dxmin$  &  $Z_{\rm disk}$ \\ 
    acronym & [$\kms$]   & [kpc]   & [kpc]     & [$\Msun$]
            & [$\Msun$]&         & [$\Msun$] &   
            & [$\Msun$]& [kpc]   & [pc]      & [$\Zsun$]  \\
    \midrule
    \dw    & $30$       & $41$    & $150$    &$10^{10}$ 
           &$3.5 \times 10^8$& $0.5$   &$3.5 \times 10^7$& $10^5$     
           & $600$      & $2.3$   & $18$      & 0.1\\
    \sbc   & $65$       & $89$    & $300$     &$10^{11}$ 
           &$3.5 \times 10^9$& $0.5$   &$3.5 \times 10^8$& $10^6$     
           & $600$      & $2.3$   & $18$      & 0.1\\
    \mw    & $140$      & $192$   & $600$       & $10^{12}$ 
           &$3.5 \times 10^{10}$& $0.3$ & $3.5 \times 10^9$& $10^6$ 
           & $10^4$     & $4.7$   & $36$       & 1 \\
    \bottomrule
  \end{tabular}
\end{table*}

We use \ramsesrt{} (\RRT{}, \RM{}), an RHD extension of the adaptive
mesh refinement (AMR) code \ramses{} \citep{Teyssier:2002fj}.
\ramses{} models the interaction of dark matter, stellar populations
and baryonic gas, via gravity, hydrodynamics and radiative
cooling. The gas evolution is computed using a second-order Godunov
scheme for the Euler equations, while trajectories of collisionless DM
and stellar particles are computed using a particle-mesh solver.
\ramsesrt{} adds the propagation of photons and their on-the-fly
interaction with hydrogen and helium via photoionisation, heating, and
momentum transfer; and with dust particles via heating and momentum
transfer. The code solves the advection of photons between grid cells
with a first order moment method and closes the set of radiation
transport equations with the M1 relation for the Eddington tensor. The
trapped/streaming photon scheme presented in \RM{} describes the
diffusion of multi-scattering IR radiation.  The radiation in a photon
\emph{group}, defined by a frequency interval, is described in each
grid cell, by the radiation energy density $E$ (energy per unit
volume) and the bulk radiation flux $\bf F$ (energy per unit area per
unit time), which corresponds approximately to the radiation intensity
integrated over all solid angles. \ramsesrt{} solves the
non-equilibrium evolution of the ionisation fractions of hydrogen and
helium, along with photon fluxes and the gas temperature in each grid
cell.

Because the timestep length, and therefore the computational load,
scales inversely with the speed of light $c$, we apply the so-called
reduced speed of light approximation \citep[][\RRT{}]{Gnedin:2001cw}
in runs that include radiation, to maintain a manageable computing
time. In this work, we use a light speed fraction $f_c=1/200$,
i.e. free-streaming photons are propagated at a speed $\cred=c/200$,
such that the timestep is most of the time limited by non-RT
conditions, and the slow-down due to RT is only about a factor 2-3
compared to HD simulations, depending on the number of photon groups
and processes included (and the inclusion of SN feedback, which limits
the timestep as well). We showed in \RRT{} that larger values for
$f_c$ than we have chosen here are preferable in simulations of galaxy
evolution in order to accurately capture the expansion speed of
ionisation fronts in the ISM, but the light speed convergence tests
presented in \App{ctest.app} indicate that our results are robust with
respect to the chosen light speed.

We run simulations of isolated rotating disk galaxies of baryonic mass
$3.5 \times (10^{8}, 10^{9}, 10^{10}) \, \Msun$ consisting of gas and
stars embedded in DM halos of masses $10^{10}$, $10^{11}$, and
$10^{12} \ \Msun$, respectively. The simulation sets, named \dw{},
\sbc{}, and \mw{}, after the order-of-magnitude of the baryonic
masses, are presented in \Tab{sims.tbl}, and the parameters listed in
the table are explained in what follows. The baryonic mass of the most
massive galaxy (\mw{}) is comparable to that of the present-day
Milky-Way (MW).

For \dw{} and \sbc{}, the host DM halos are disproportionally low in
mass, compared to results from abundance-matching
\citep{Moster:2013dl} and cosmological simulations that match the
observed galaxy mass function \citep{Schaye:2015gk}. These
under-massive DM halos are not a major issue for the current work,
however. We are primarily interested in comparing the relative effects
of different feedback processes on the properties of the galaxy disk,
for which the dark matter profile does not play an important role. To
verify that our results are insensitive to the mass of the host halo,
we have run counterparts of the least massive galaxy, \dw{}, with the
halo mass increased to a more realistic value
$\Mhalo=7 \times 10^{10} \ \Msun$ (i.e. an increase by a factor of
seven compared to \Tab{sims.tbl}), while keeping the same
resolution. We confirmed that while the simulations were more
expensive due to the increased size of the box and number of DM
particles, the results were not affected.

\subsection{Initial conditions}
The initial conditions (ICs) are generated with the {\sc
  MakeDisk}\footnote{Adapted to generate \ramses{}-readable format by
  Romain Teyssier and Damien Chapon.} code by Volker Springel
\citep[see][]{Springel:2005co,Kim:2014dj}. The DM halos follow an NFW
density profile \citep{Navarro:1997if} with concentration parameter
$c=10$ and spin parameter $\lambda=0.04$. We model the dark matter in
each halo with $\Npart$ collisionless particles of identical mass. The
initial disk consists of gas cells and $\Npart$ identical mass stellar
particles, both set up with density profiles that are exponential in
radius and Gaussian in height above the mid-plane. The galaxies also
contain stellar bulges with mass one tenth of the stellar disk mass,
represented by $0.1 \, \Npart$ particles. The stellar particles that
are present at the beginning of the simulation do not perform any
feedback. The initial gas profiles do not enforce exact hydrostatic
equilibrium. However, the initial (few million years) stabilisation of
the galaxy, which manifests itself in contraction of the inner dense
gas and expansion of the outer diffuse gas, is minor, as can be
inferred from plots of the star formation rate
(e.g. \Fig{sfr_G9.fig}). The initial temperature of the gas disk is
$T=10^4$ K, and the disk metallicity, $Z_{\rm disk}$, is set to a
constant value, either $0.1$ or $1$ times Solar (see \Tab{sims.tbl}),
with the metal mass fraction in the Sun taken to be $\Zsun=0.02$. The
circumgalactic medium (CGM) initially consists of a homogeneous hot
and diffuse gas, with $\nh=10^{-6} \ \cci$, $T=10^6 $ K and zero
metallicity. The cutoffs for the disk's radial and vertical gas
profiles, which mark the transition between the disk and CGM, are
chosen to minimize the density contrast between the disk edges and the
CGM.

\subsection{Star formation}
Star formation follows a standard Schmidt law. In each cell where the
gas density exceeds the chosen star formation threshold
\begin{align}
\ns = 10 \ \cci,
\end{align}
gas is converted into stars at a rate 
\begin{align} \label{SFR.eq}
\dot \rho_{*} = \sfeff \rho / t_{\rm ff}, 
\end{align}
where $\rho$ is the gas density and $\sfeff=0.02$ is the star
formation efficiency per free fall time,
$t_{\rm ff} = \left[ 3 \pi/(32 G \rho) \right]^{1/2}$, where $G$ is
the gravitational constant.  Collisionless particles of mass $m_*$,
representing stellar populations, are formed stochastically from the
gas, with the probability of forming one drawn from a Poissonian
distribution \citep[for details see][]{Rasera:2006gz}.  \Tab{sims.tbl}
lists the stellar particle masses used in the simulations. In addition
to the density threshold for star formation, we also do not allow
stars to form in gas warmer than $T/\mu=3000$ K, where $\mu$ is the
average particle mass in units of the proton mass. We note, however,
that our results are insensitive to increasing or even removing the
temperature threshold.

\subsection{Supernova feedback}
We model SN feedback with a single injection from each stellar
particle into its host cell, $5$ Myr after the particle's birth, of
mass $m_{\rm ej}=\eta_{\rm SN} \times m_*$, and thermal energy
$\epsilon_{\rm SN}=\eta_{\rm SN} \times 10^{51} \ {\rm{erg}} \ m_*/10
\ \Msun$.
We use $\eta_{\rm SN}=0.2$, roughly corresponding to a
\cite{Chabrier:2003kia} stellar initial mass function (IMF). We
neglect the metal yield associated with stellar populations, i.e. the
stellar particles inject zero metals into the gas.

At our resolution, the ``thermal dump'' SN feedback model that we use
is known to suffer from numerical overcooling
\citep[e.g.][]{Creasey:2011ef, DallaVecchia:2012br, Creasey:2013gu},
but we use it here, because it is simple and because it allows us to
investigate how far radiation feedback can go to compensate for its
low efficiency. The coupling between radiation and SN feedback, which
we study in \Sec{SN_boost.sec}, could depend on the choice of SN
feedback model. More efficient SN feedback might either be amplified
more efficiently by the stellar radiation to suppress star formation
and increase outflow rates, or conversely, it might dominate
completely over the effects of radiation feedback and render it
negligible. These considerations are beyond the scope of the present
paper, but in future work we will combine radiation feedback with more
efficient recipes for SN feedback, to find what combination produces
best agreement with observations (which is not the point of this
paper) and to study how the interplay of the feedback processes is
affected.

\subsection{Gas thermochemistry}
We evolve the thermochemistry semi-implicitly with the method
presented in \RRT{}. The method tracks the non-equilibrium cooling
rates of hydrogen and helium, here assuming zero incoming photon
flux. The ionisation fractions of hydrogen and helium are stored in
each cell as three passive scalars, which are advected with the
gas. We assume hydrogen and helium mass fractions $X=0.76$ and
$Y=0.24$, respectively, and Solar ratios for the metal species,
i.e. we track a single scalar representing the metal mass fraction in
each cell.

We add the contribution from metals to the cooling rate using tables
generated with {\sc cloudy} \citep{Ferland:1998ic}, assuming
photoionisation equilibrium with the redshift zero
\cite{Haardt:1996fq} UV background. With metal cooling, the gas can in
principle cool non-adiabatically to $\sim 10$ K.  We do not model the
change in the metal cooling rate with the local radiation flux, which
may affect galaxy evolution
\citep[e.g.][]{Cantalupo:2010jw,Kannan:2014di}. In future work, we
will consider more realistic metal cooling, which takes the local
radiation flux into account.

\subsection{Adaptive refinement}
In the adaptive refinement scheme of \ramses{}, cells can be split
into 8 child cells of width half that of the parent. The width of a
cell is determined by its refinement hierarchy level $\ell$, by
$\dx_{\ell}=\Lbox/2^{\ell}$, where $\Lbox$ is the simulation box
width. The maximum and minimum cell widths, $\dx_{\rm max}$ and
$\dx_{\rm min}$, are determined by the enforced minimum and maximum
allowed refinement levels in a simulation, which in this work are
$\dx_{\rm max}= 2-5$ kpc and $\dx_{\rm min}= 18-36$ pc, depending on
the simulation set (see \Tab{sims.tbl}). Adaptive refinement follows
mass: a cell is refined if it contains 8 or more collisionless
particles, if the cell gas mass $\mcell > 12 \ \mstar$, or if $\dx$ is
more than a quarter of the local Jeans length.
 
\subsection{Artificial ``Jeans pressure''}
We impose a pressure floor on gas to prevent artificial fragmentation
below the Jeans scale \citep{Truelove:1997bj}.  The Jeans length scale
for a self-gravitating cloud is 
\begin{align} \label{LJeans.eq}
  \LJeans = \sqrt{\frac{\pi \csound}{G \rho}}
  = 16 \ {\rm pc} \ \left( \frac{T}{1 \ {\rm K}} \right)^{1/2}
  \ \left( \frac{\nh}{1 \ \cci} \right)^{-1/2},
\end{align}
where $\csound=\sqrt{\gamma \kb T / \mp}$ is the sound speed, and we
assumed a ratio of specific heats of $\gamma=1.4$, appropriate for a
monatomic gas. From \Eeq{LJeans.eq}, the requirement that the Jeans
length is resolved by at least N cell widths becomes a temperature
floor of the form
\begin{align} \label{Tfloor.eq}
 \frac{T}{1 \ {\rm K}} \ge
  \frac{\nh}{1 \ \cci} \ \left( \frac{N \dx}{16 \ {\rm pc}} \right)^2
\end{align}
We apply this floor in the form of an effective temperature function,
\begin{align} \label{polytrope.eq}
  T_{\rm J} = T_0 \ \nh / \ns,
\end{align}
where we use $T_0=500$ K in all our simulations, ensuring that the
Jeans length is resolved by a minimum number of 6 cell widths in \dw{}
and \sbc{} and 3 cell widths in \mw{}. The pressure floor is
non-thermal, and added to the physical temperature, $T$, and hence we
can have $T \ll T_{\rm J}$.

\subsection{Radiation feedback}
We include the emission and propagation of stellar radiation, and its
interaction with the gas. The mass-, age- and metallicity-dependent
stellar specific luminosities are extracted on the fly from the
spectral energy distribution (SED) model of \cite{Bruzual:2003ck}, as
described in \RRT{}, assuming a \cite{Chabrier:2003kia} IMF. Stellar
particles inject photons into their host grid cells at every fine RHD
timestep.

\begin{table*}
  \begin{center}
  \caption
  {Photon group energy (frequency) intervals and properties. The
    energy intervals defined by the groups are indicated in units of
    eV by $\epsilon_0$ and $\epsilon_1$ (in units of \AA{}ngstrom by
    $\lambda_0$ and $\lambda_1$). The next four columns show photon
    properties derived every $5$ coarse time-steps from the stellar
    luminosity weighted SED model (see \Fig{groups.fig} and
    surrounding text). These properties evolve over time as the
    stellar populations age, and the approximate variation is
    indicated in the column headers. $\eavg$ denote the photon
    energies, while $\csavg_{\hi}$, $\csavg_{\hei}$, and
    $\csavg_{\heii}$ denote the cross sections for ionisation of
    hydrogen and helium, respectively. $\kO$ is the dust opacity. The
    gas opacity scales with the gas metallicity,
    $\ki=\kOi \, Z/\Zsun$, where $i$ denotes the photon group.}
  \label{groups.tbl}
  \begin{tabular}{l|R{2}{2}R{2}{2}rr|rrrr|r}
    \toprule
    Photon & \multicolumn{1}{c}{$\epsilon_0$ [eV]} 
    & \multicolumn{1}{c}{$\epsilon_1$ [eV]}
    & $\lambda_0$ [\AA{}] & $\lambda_1$ [\AA{}]
    & $\eavg$ [eV] & $\csavg_{\hi} \, [\cs]$ 
    & $\csavg_{\hei} \, [\cs]$ & $\csavg_{\heii} \, [\cs]$
    & $\kO \, [\ccg]$ \\
    group & \multicolumn{1}{c}{} 
    & \multicolumn{1}{c}{}
    &  & 
    & $\pm 10 \%$ & $\pm 5 \%$ 
    & $\pm 5 \%$ & $\pm 5 \%$
    &  \\
    \midrule
    IR      & $0.1$  & $1$    & $1.2 \times 10^5$ & $1.2 \times 10^4$ 
            & $0.6$ & 0     & 0 & 0 & $10$ \\
    Opt     & $1$    & $13.6$ &  $1.2 \times 10^4$ & $9.1 \times 10^{2}$ 
            & $5.5$  & 0 & 0 & $0$ & $10^3$ \\
    UV$_{\hi}$ & 13.6 & 24.59 & $9.1 \times 10^{2}$ & $5.0 \times 10^{2}$ & 18.0 
            & $3.3 \times 10^{-18}$ & 0 & 0 & $10^3$ \\
    UV$_{\hei}$    & 24.59 & 54.42 & $5.0 \times 10^{2}$ 
            & $2.3 \times 10^{2}$ & 33.4 
            & $6.3 \times 10^{-19}$& $4.8 \times 10^{-18}$ & 0 & $10^3$ \\
    UV$_{\heii}$   & 54.42  & \multicolumn{1}{c}{$\infty$} 
            & $2.3 \times 10^{2}$ & 0
            & 60.0 & $9.9 \times 10^{-20}$ & $1.4 \times 10^{-18}$ & 
                                        $1.3 \times 10^{-18}$    & $10^3$ \\
    \bottomrule
  \end{tabular}
  \end{center}
\end{table*}

We bin the radiation into five photon groups, defined by the photon
energy intervals listed in \Tab{groups.tbl}. The groups are, in order
of increasing energy, IR, optical, and three groups of ionising
ultraviolet (UV) photons, bracketed by the ionisation energies for
\hi{}, \hei{}, and \heii{}. We include the first two groups only in
runs with radiation-dust interactions, while we include the three UV
groups in all runs with radiation. \App{groups.app} describes how the
stellar luminosities, photon group energies, and ionisation cross
sections are derived from the SED model. \Tab{groups.tbl} lists
typical values for the energies and cross sections, along with their
variations over the simulation run-time.

An important advantage of the moment method that we use for the
radiative transfer is that the computational cost, i.e. the runtime of
the simulations, is \emph{independent} of the number of radiation
sources. With the alternative class of ray-tracing methods
\citep[e.g.][]{Wise:2011iw}, the computational cost increases more or
less linearly with the number of sources, which requires remedies to
keep down the computing cost, such as merging of sources or rays
\citep[e.g.][]{Pawlik:2008kk} and/or turning them off after a few
Myrs. Turning them off seems acceptable, considering that the
luminosity of a stellar population has dimmed by orders of magnitude
$10$ Myrs after its birth (see \Fig{groups.fig}). However,
\cite{Kannan:2014di} have pointed out that the cumulative radiation
from many such dim old sources may play a role in stellar
feedback. Since we do not have an issue with the number of radiation
sources in our implementation, stellar particles are never turned off
after their birth, and the cumulative radiation from old populations
is included.

We implement three ``separate'' radiation feedback processes,
describing different interactions between the radiation and gas:
\begin{enumerate}
\item Photons ionise and heat the gas they interact with, following
  the thermochemistry described in \RRT{}, typically heating the
  ionised gas to $\approx 2 \times 10^4$ K. All our runs with
  radiation include this process. We describe in \App{groups.app} how
  we use the SED model from \cite{Bruzual:2003ck} to derive
  photoionisation cross sections, which are typically a few times
  $10^{-18} \ \cs$ for \hi{}, \hei{}, and \heii{}.
\item Direct pressure, i.e. momentum transfer, from the ionising
  photons onto the gas.
\item Indirect radiation pressure on the gas, via dust particles, from
  the ionising photons, optical photons, and from reprocessed IR
  radiation, where the latter multi-scatters.
\end{enumerate}
\RM{} contains a detailed description of the implementation and tests
of the latter two processes, including the diffusion, pressure, and
work of multi-scattered IR radiation. We perform the correct diffusion
of IR radiation by a partition in every cell into free-streaming and
trapped photons, where the trapped photons dominate in the case of
large optical depth on the scale of the cell width.

We will refer to the various radiation feedback processes under the
collective acronym of RT (radiative transfer) feedback. RT feedback
may thus refer to the inclusion of any or all of the radiation
feedback processes under consideration. We will successively add the
three RT processes to our simulations, to probe their respective
importance. Always included in RT feedback is photoionisation and
photoionisation heating, from the three UV photon groups. Onto that we
add direct pressure from photoionisation (again only the three UV
groups). Finally, we add radiation-dust interactions, from all five
groups.

Each photon group $i$ has a dust-interaction opacity, $\kOi$, listed
in the rightmost column of \Tab{groups.tbl}. The gas absorbs momentum
from the photons (via dust) with the gas opacity
\begin{align}
\ki=\kOi \, Z / \Zsun,
\end{align}
i.e. in our model, the dust content simply scales with the metallicity
of the gas. Higher energy photons (all but IR) absorbed by dust are
reprocessed, i.e. re-emitted, into the IR group, while IR photons are
(multi-) scattered by the dust.

For the IR we assume an opacity of $\kIR= 10 \, Z / \Zsun \, \ccg$,
while for the higher energy photons we assume
$\kUV= 10^3 \, Z / \Zsun \, \ccg$, i.e. a hundred times higher than
that of the IR. These opacities are physically motivated from a
combination of observations and dust-formation theory of the ISM and
stellar nurseries (\citealp{Semenov:2003hk} for IR,
\citealp{Li:2001te} for higher energy radiation), but they are
uncertain by a factor of few, due to model uncertainties and the
temperature dependency, which we ignore. Similar values have been used
in e.g. \cite{Hopkins:2012hr}, \cite{Agertz:2013il}, and
\cite{Roskar:2014be}. The IR opacity we use is at the high-end of what
is usually considered in the literature, which is
$\kOIR \approx 5-10 \ \ccg$. We have tested and confirmed that our
results are insensitive to order-of-magnitude variations in the dust
opacities (see \Sec{whatif.sec}).

There are two important exceptions from the default behaviour of the
implementation described in \RM{}. 

Firstly, our resolution of $\sim 10$ pc does not allow us to accurately
capture the regime where dust is optically thick to photons, and
radiation and dust are coupled via absorption and blackbody
emission. For this reason, and also for the sake of simplicity, we
exclude the dust temperature evolution (\S 2.3.2 in \RM{}), where the
gas temperature is coupled directly to the IR radiation temperature
via the Planck cross section. We decouple the dust temperature by
simply setting the Planck cross section to zero (while keeping a
nonzero Rosseland opacity).

\begin{figure*}
  \centering
  \includegraphics[width=\textwidth]
    {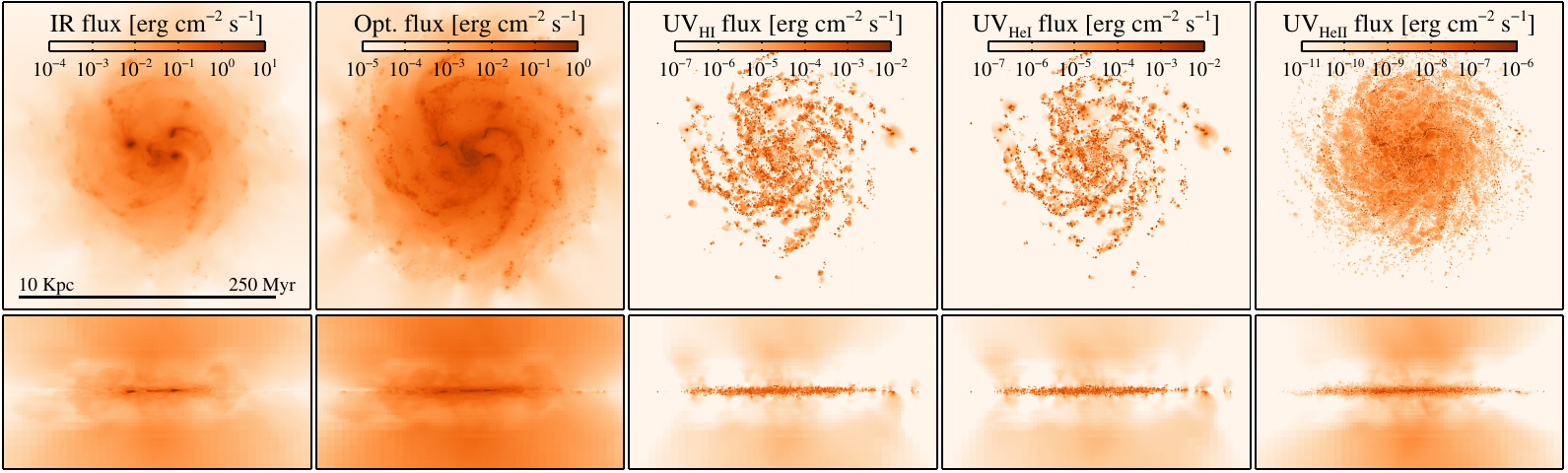}
  \caption
  {\label{G9_photons.fig}Illustration of radiation flux in the five
    photon groups included in this work. The maps show
    density-weighted solid-angle integrated photon fluxes, $\cred E$,
    along the LOS in the \sbc{} galaxy with SN and full RT feedback
    (\sbc\_\sime{}) at $250$ Myr. The photon groups are shown by
    increasing photon energy, from left to right. The upper row shows
    the galaxy face-on and the lower row shows it edge on. The much
    larger contrast in the fluxes of the ionising photons (three
    right-most panels), owes to their much shorter mean free
    paths. Also, to a smaller degree, the optical photons have larger
    contrast than IR radiation, for the same reason. For the
    corresponding distribution of stars and gas in the same snapshots,
    this figure can be compared to \Fig{maps_G9.fig} (bottom left
    panel).}
\end{figure*}

The second change is that we assume a fully directional radiation in
each cell for the free-streaming radiation pressure (\RM{}, Eq. 28),
by using a renormalised radiation flux magnitude of $\cred E$, rather
than the actual radiation flux of $|{\bf F}| \le \cred E$. We do this
to counter a resolution effect, as the reduced flux,
$\fred\equiv |{\bf F}|/\cred E \le 1$, takes a few ($\sim 5$) cell
widths to evolve to unity with our advection scheme, even with
free-streaming radiation. We demonstrate this numerical effect with a
simple idealised experiment in \App{redflux.app}. For the cell
containing the emitting source, this resolution artefact is obvious,
since the radiation is isotropic and hence has zero bulk flux (only
$E$ is incremented with stellar emission). The lack of bulk radiation
flux very close to the emitting stellar particles diminishes the
effect of radiation pressure, especially since it turns out that HII
regions are often poorly resolved in our simulations. Therefore, we
apply this \emph{full reduced flux approximation} ($\fred=1$) for the
radiation pressure, to compensate for resolution effects. It can then
be argued that we overestimate radiation pressure, especially in
regions where cancellation effects are relevant, but since it turns
out that radiation pressure is very weak in our simulations, we prefer
to be in danger of overestimating rather than the opposite. We do not
apply the full reduced flux approximation for the IR photon group,
since pressure from the IR radiation, in the limit where the optical
depth is not resolved, is accurately captured by the radiation
trapping scheme (\RM{}).

\Fig{G9_photons.fig} illustrates the distribution of photons, for the
five radiation groups, in one of our runs of the intermediate mass
galaxy disk (\sbc{}). The figure shows mass-weighted averages along
lines-of-sight (LOS) of photon fluxes, integrated over all solid
angles, i.e. the mapped quantity is $\cred E$, where $\cred$ is the
reduced speed of light and $E$ is the radiation energy density. From
left to right, the maps show photon groups with increasing energy,
starting with IR on the far left, the optical, and finally the three
ionising groups.  The photon fluxes differ greatly between the photon
groups, decreasing with increasing photon energy. We use different
color scales, such that the logarithmic range is the same, but the
upper limit roughly matches the maximum flux in each set of
face-on/edge-on maps. For the highest energy group (far right) the low
luminosity is simply due to the low emissivity from the stellar
populations (see \Fig{groups.fig}, where we plot the emissivity of the
stellar populations). For the two lower energy ionising groups (second
and third from right) the stellar emissivity is similar to that of the
optical group, yet the galaxy luminosity is clearly much lower than in
the optical. This is due to the much more efficient absorption of the
ionising photons. For photoionisation of hydrogen and helium, the
opacities are $\sigma/m_p \sim 6 \times 10^{5} \, \ccg$, where
$\sigma \sim 10^{-18} \, \cs$ is the photoionisation cross section
(see \Tab{groups.tbl} and \App{groups.app}) and $m_p$ is the proton
mass, while for the optical group the opacity is
$\kOpt=\kOOpt Z/\Zsun = 10^2 \, \ccg$.  Hence the difference in
opacities is more than three orders of magnitude. While the ionising
photons are absorbed close to their emitting sources, the optical
photons are much more free to propagate through the disk and escape
from it. The direct stellar IR emission is relatively dim, about three
orders of magnitude lower than that of the optical group, yet the maps
on the far left show that the radiation energy flux is highest in the
IR group. This is because the IR photons are mostly reprocessed from
the Optical and UV photons, which are captured by the dust and
re-emitted into the IR.

\begin{figure}
  \centering
  \subfloat
  {\includegraphics[width=0.22\textwidth]
    {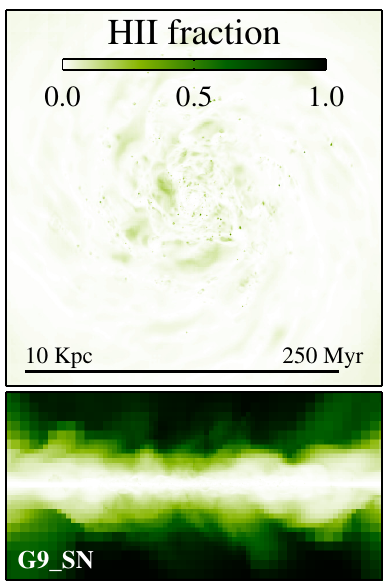}}
  \subfloat
  {\includegraphics[width=0.22\textwidth]
    {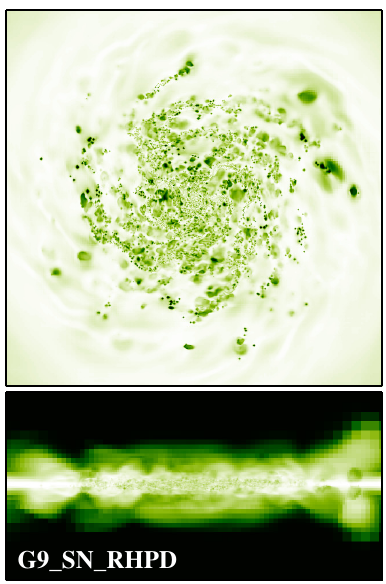}}
  \caption
  {\label{maps_G9_xHII.fig}Ionised hydrogen fractions in the \sbc{}
    galaxy at 250 Myr. The maps show mass-weighted ionised fractions
    along the LOS, for SN feedback only (left, \sbc\_\simb{}) and
    added (full) radiation feedback (right, \sbc\_\sime{}). The right
    map is the same snapshot as shown in the panels of
    \Fig{G9_photons.fig}.}
\end{figure}

\Fig{maps_G9_xHII.fig} illustrates the effect of the radiation on the
hydrogen ionisation fractions in the gas, which are tracked by the
code. The left panel shows a run with SN feedback only, while the
right panel shows the same galaxy also with full RT feedback, which
results in an abundance of dense photoionisation-powered \hii{}
regions.
 
\subsection{Overview}

\Tab{sims.tbl} lists the properties of the simulated galaxies.  We run
each simulation for $500$ Myr. \Tab{fb.tbl} lists the $6$ combinations
of four feedback processes included in the simulations: No feedback at
all (\sima), SN feedback only (\simb), with added radiation feedback
with radiation heating only (\simc), with added direct pressure from
ionising photons (\simd), with added radiation pressure on dust, and
optical and (reprocessed) IR radiation groups (\sime), and, finally,
with all radiation feedback processes included, but without SN
feedback (\simf). The name of each run is a combination of the
acronyms from Tables \ref{sims.tbl} and \ref{fb.tbl}, e.g. the name
\sbc\_\simd{} represents the \sbc{} galaxy (baryonic mass of
$3.5 \times 10^{9} \ \Msun$), simulated with SN feedback and ionising
stellar radiation with heating and direct pressure.

\begin{table}
  \centering
  \caption
  {Feedback processes included in the simulations}
  \label{fb.tbl}
  \begin{tabular}{l|cccc}
    \toprule
    Feedback& SN       & Radiation  & Direct rad. & Dust    \\ 
    acronym & feedback & heating    & pressure    & pressure \\
    \midrule
    \sima    &         &           &        &        \\
    \simb    &  \cmark &           &        &        \\
    \simc    &  \cmark & \cmark    &        &        \\
    \simd    &  \cmark & \cmark    & \cmark &        \\
    \sime    &  \cmark & \cmark    & \cmark & \cmark \\
    \simf    &         & \cmark    & \cmark & \cmark \\
    \bottomrule
  \end{tabular}
\end{table}

\section{Results} \label{results.sec} We now turn to the simulated
galaxy disks and examine the impact of radiation feedback and the
interplay of radiation and SN feedback. We start with the
intermediate-mass galaxy, which has the highest resolution in terms of
the number of volume/particle elements, and then consider in turn,
with somewhat less detail, the more massive \mw{} galaxy and the less
massive \dw{} galaxy.

\subsection{G9: intermediate mass gas-rich galaxy}
\begin{figure*}
  \centering
  \subfloat
  {\includegraphics[width=0.494\textwidth]
    {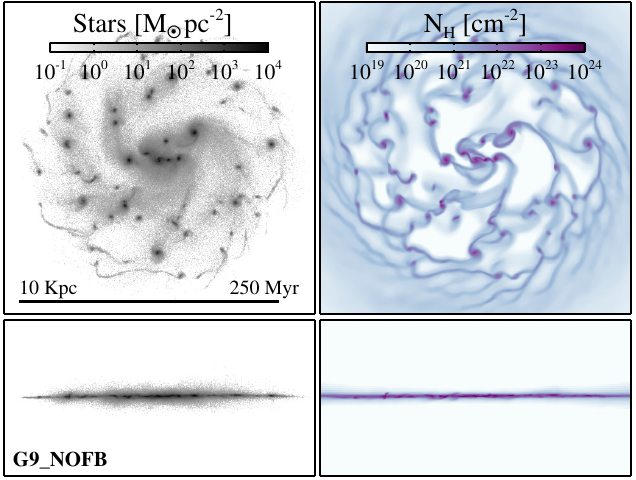}}
  \hspace{0.mm}
  \subfloat
  {\includegraphics[width=0.494\textwidth]
    {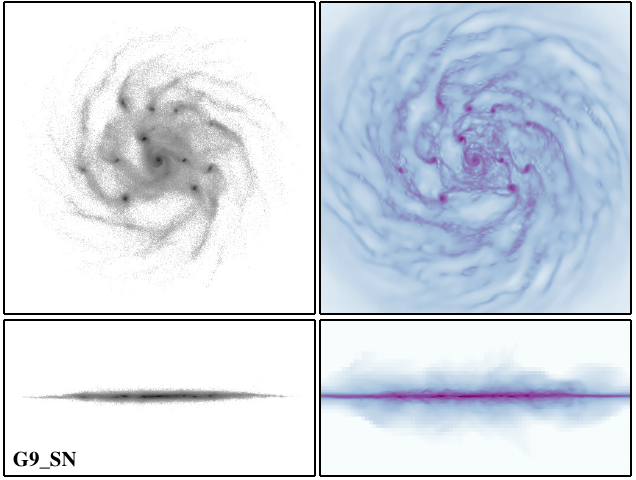}}
  \vspace{0.5mm}
  \subfloat
  {\includegraphics[width=0.494\textwidth]
    {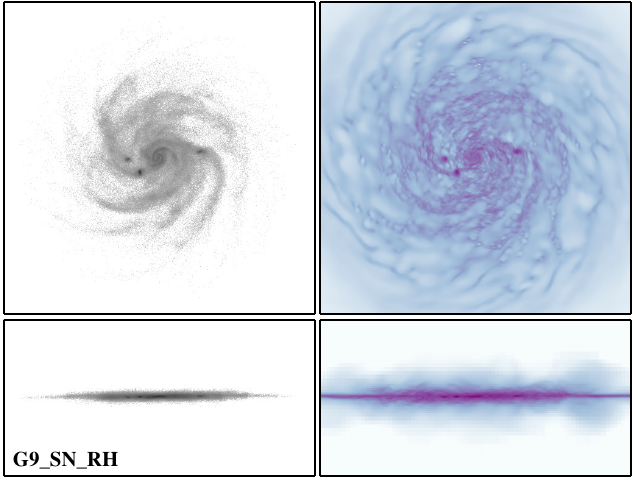}}
  \hspace{0.0mm}
  \subfloat
  {\includegraphics[width=0.494\textwidth]
    {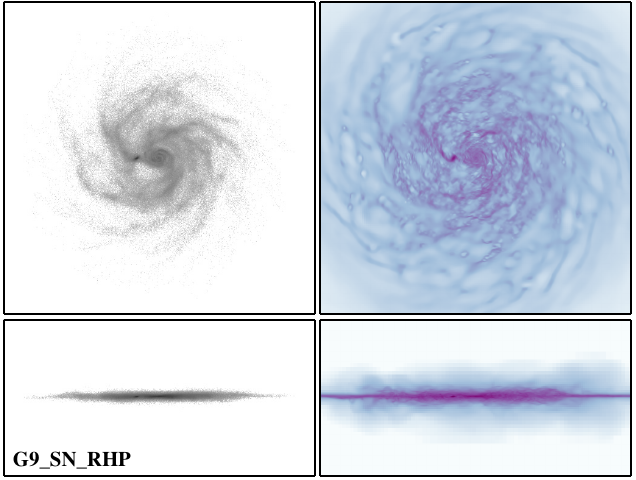}}
  \vspace{0.5mm}
  \subfloat
  {\includegraphics[width=0.494\textwidth]
    {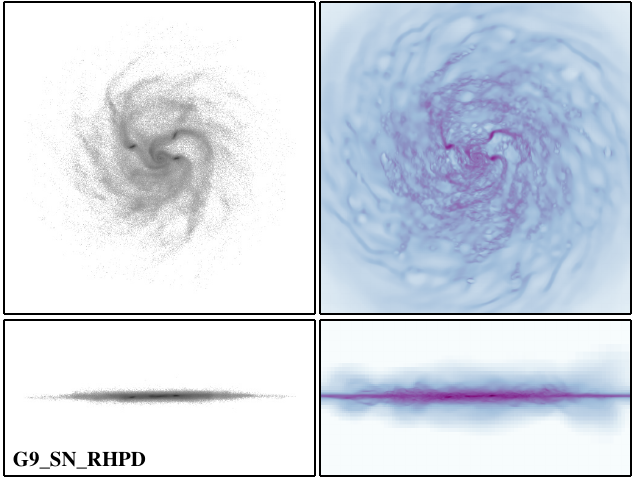}}
  \hspace{0.mm}
  \subfloat
  {\includegraphics[width=0.494\textwidth]
    {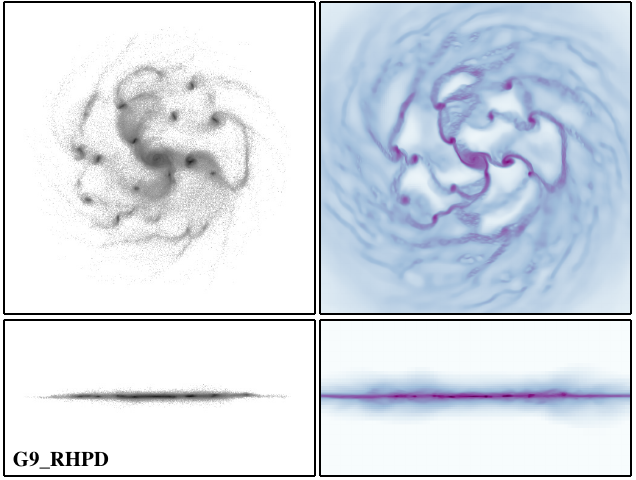}}
  \caption
  {\label{maps_G9.fig}Maps of the \sbc{} galaxy (roughly ten times
    less massive in baryons than the MW) at $250$ Myr, for the
    different feedback runs. Each panel shows face-on and edge-on
    views of the stellar density (left) and the hydrogen column
    density (right). From left to right, top to bottom, the panels
    show the runs without feedback (\sima{}), SN feedback only
    (\simb{}), SN and radiation heating (\simc{}), SN+radiation
    heating + direct pressure (\simd{}), SN + radiation heating +
    direct + dust pressure (\sime{}), and radiation heating + direct +
    dust pressure (\simf{}). The physical length scale and the color
    scales for the stellar and gas column densities are shown in the
    top left panel. The addition of radiation feedback smooths and
    thickens the disk, compared to SN feedback only. The respective
    additions of direct UV radiation pressure and then optical and IR
    pressure have little effect.}
\end{figure*}

We first focus on the intermediate mass galaxy, \sbc{} ($\approx$
one-tenth the baryonic mass of the MW), and we begin by considering
the qualitative effects of the different feedback processes on the
morphology of the disk. \Fig{maps_G9.fig} shows maps of stellar
density and total hydrogen column density, face on and edge on, at
$250$ Myr, which is half the run duration. Without feedback (top left
panel), the galaxy contains many cold star-forming clumps
interconnected by narrow gas filaments. SN feedback (top right panel)
dramatically reduces star and clump formation, especially at large
radii, smooths out the gas distribution and thickens the gas disk
compared to the no feedback case. The inner $\sim 3$ kpc from the
center of the galaxy remain quite clumpy, however. The addition of
ionising radiation and photoionisation heating (middle left panel)
adds to the effect of SN feedback by further smoothing the morphology
of the galaxy, and further reducing the number of clumps. The addition
of radiation pressure, direct (middle right) and on dust (bottom
left), has little impact. With SN feedback excluded, radiation heating
and pressure on its own (bottom right) is insufficient to prevent
massive clump formation in the galaxy, and it is noticeably more
clumpy and thinner than with SN feedback only.

\subsubsection{Star formation}

\begin{figure}
  \centering
  \includegraphics
    {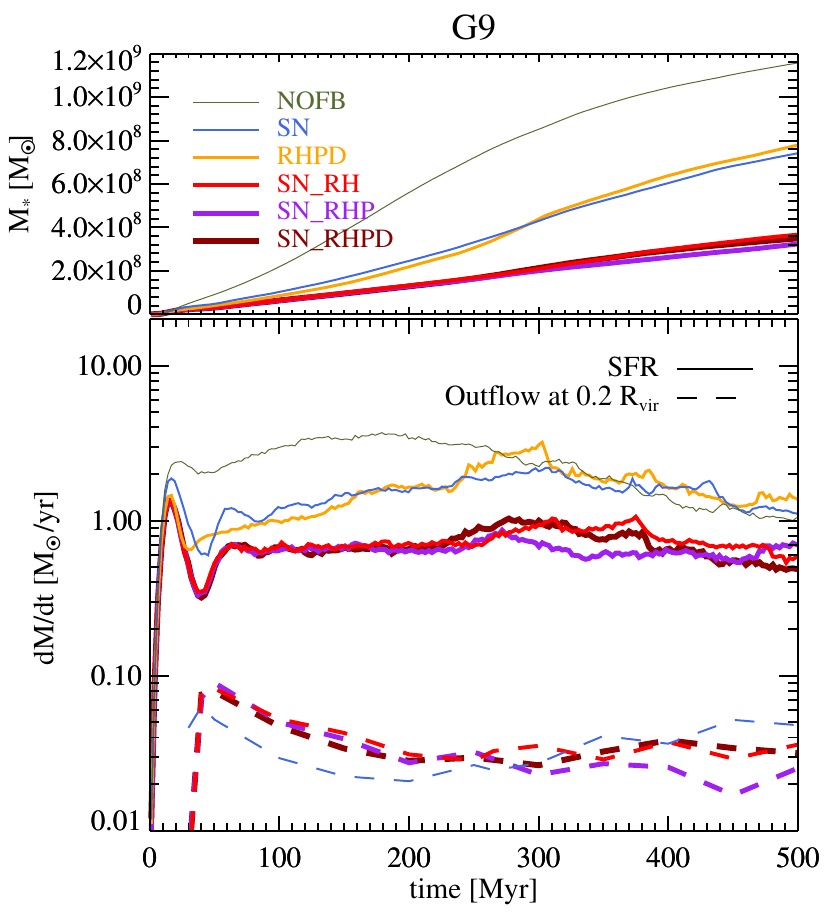}
  \caption
  {\label{sfr_G9.fig}Star formation and outflows in the \sbc{} runs
    with different feedback processes included, as indicated in the
    legend: no feedback (\sima{}), SN feedback only (\simb{}), RT
    feedback only with all processes activated (\simc{}), SN+RT
    feedback with RT heating only (\simd{}), the same with added
    direct ionising radiation pressure (\sime{}), and then with added
    lower energy radiation and dust pressure (\simf{}). {\bf Upper
      panel:} stellar mass formed over time. {\bf Lower panel:} star
    formation rates (solid lines) and outflows across planes at
    distances of $0.2 \ \Rvir$ from the disk plane (dashed
    lines).}
\end{figure}

Star formation is the most direct probe of the efficiency of feedback
processes. The more efficient the feedback, the more it will reduce
and regulate star formation. \Fig{sfr_G9.fig} shows, for the \sbc{}
galaxy, the cumulative stellar mass formed over time (upper panel) and
star formation rates (solid lines in the lower panel). These results
are in line with the qualitative effects we saw in the previous
maps. Compared to the no feedback case, turning on SN feedback reduces
the formation of stars by about $35\%$ at $500$ Myr. Turning instead
to radiation feedback, with both the pressure terms included, gives a
very similar reduction in the star formation. Combining SN and full
radiation feedback (three thickest curves) considerably reduces the
star formation again, by $\approx 70\%$ compared to the no feedback
case, and by $\approx 50\%$ compared to the cases with SN or radiation
feedback only.

We can probe the importance of radiation pressure by comparing the
curves where SN feedback is combined with successive introductions of
radiation feedback processes, i.e. photoionisation heating, direct
ionising radiation pressure, and radiation pressure on dust. The
stellar mass formed is very similar, indicating that radiation heating
is the dominant radiation feedback process.  

The $\la 10\%$ variation in the stellar mass formed at the end of the
runs for the various radiation feedback processes, is too slight to
require serious interpretation. It is likely a random effect where
small variations in the feedback model trigger massive clump formation
at different times in the simulations. Individual clumps can dominate
the star formation for tens of Myrs, while they migrate to the center
of the disk.  While these clump formations are likely random, we
cannot rule out the possibility that these effects of successively
added radiation feedback processes are systematic. If the effect is
real and non-stochastic, the way it works is somewhat
counter-intuitive, as the addition of radiation pressure on dust and
the sub-ionising photon groups (\sime{}) \emph{boosts} star
formation. This implies \emph{negative feedback}, which can be
explained by a scenario where radiation pressure sweeps the gas into
concentrated star-forming shells or clumps. However, we do not see a
negative feedback effect from radiation pressure on dust in the other
galaxies considered in this paper, and hence we conclude that it is a
random effect, rather than systematic.

Focusing on the star formation rates for the different feedback
processes, in the bottom panel of \Fig{sfr_G9.fig}, we see that the
combined SN+RT feedback flattens out the star formation history
compared to the case of no feedback or individual SN or RT
feedback. The star formation rates decline in the latter half of the
runs with no feedback or individual SN or RT feedback. This is due to
the galaxy disk starting to be starved of gas. The initial disk gas
mass is $\approx 2 \times 10^9 \, \Msun$, and it is clear from the
upper panel that a considerable fraction of this mass has already been
converted into stars at $500$ Myr. This narrows the difference in star
formation rates between the runs: while the rate is suppressed by a
factor $\approx 2$ at $500$ Myr by the combination of SN and RT
feedback, and not suppressed at all by only SN or RT feedback, the
suppression factor is much higher before gas depletion sets in,
peaking at a factor $\approx 5$ at $\approx 150$ Myr for the combined
feedback case and a factor $2-3$ for the ``single'' feedback case
(excluding the first $\approx 50$ Myr, when the disc is relaxing).

\subsubsection{Outflows}
\begin{figure}
  \centering
  \subfloat
  {\includegraphics[width=0.24\textwidth]
    {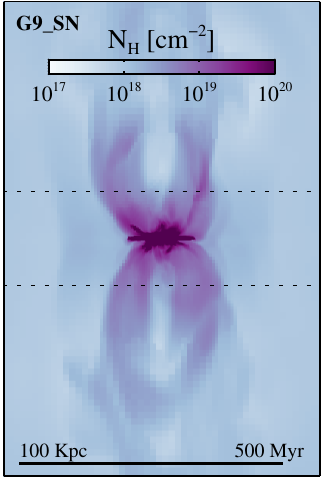}}
  \subfloat
  {\includegraphics[width=0.24\textwidth]
    {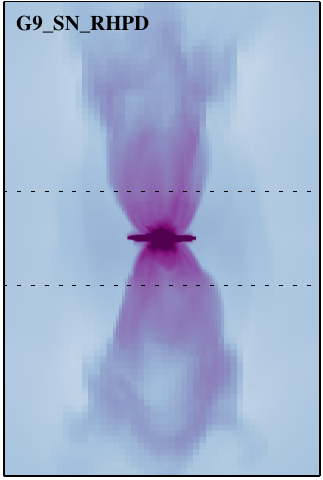}}
  \caption
  {\label{maps_G9_outflows.fig}Outflows from the \sbc{} galaxy at 500
    Myr. The maps show total hydrogen surface density for SN feedback
    only (left) and added (full) radiation feedback (right). The time,
    color-, and length-scales are marked in the left map. Dotted
    horizontal lines mark planes $0.2 \, \Rvir$ from the galaxy plane,
    where we measure the outflow/inflow rates plotted in
    \Fig{sfr_G9.fig}.}
\end{figure}

Galaxies produce outflows, and it has been suggested that radiation
feedback, and radiation pressure in particular, may be important for
generating these galactic winds \citep{Murray:2011en}.
\Fig{maps_G9_outflows.fig} shows edge-on maps of the total hydrogen
column density for the \sbc{} galaxy at $500$ Myr, with SN feedback
only (left) and with added full RT feedback (right). The panels show
that winds are generated in the \sbc{} galaxy. The winds are produced
by SN feedback: maps (not shown) with no feedback or RT feedback only
show no sign of winds. The figure reveals slightly different wind
morphologies, with the \sime{} case showing a more collimated wind
than the \simb{} run, where the wind seems to form a conical shell,
i.e. with a gas-free zone along the z-axis through the center of the
disk. This difference is due to the star formation being more
concentrated towards the center of the disk in the SN+RT feedback
case, while it is located in a few clumps at various radii from the
center in the SN case.

We consider the winds more quantitatively in the dashed curves in the
lower panel of \Fig{sfr_G9.fig}, which show gas outflow rates across
disk-parallel planes at $|z|=17.8$ kpc, or $0.2 \, \Rvir$, in each
direction from the disk. The planes are indicated by dashed horizontal
lines in \Fig{maps_G9_outflows.fig}. These are gross outflow rates,
i.e. we exclude from the calculation those cells intersecting the
planes that have inflowing gas velocity.  Where outflows exist across
those planes, which is in all the runs with SN feedback included, the
outflow rates are similar, within roughly a factor of two of each
other. RT feedback has very little effect on the outflow rates,
regardless of whether or not radiation pressure is included.

Outflows are often quantified in terms of the mass loading factor,
which is the ratio between the outflow rate and the star formation
rate. In the case of \Fig{sfr_G9.fig}, the mass loading is quite low,
i.e. the outflow rates are more than an order of magnitude less than
the star formation rates. Although the outflow rates change little
with the addition of radiation feedback, the mass loading is typically
a few tens of percent higher, since the star formation is less
efficient.

\begin{figure}
  \centering
  \includegraphics[width=0.45\textwidth]
    {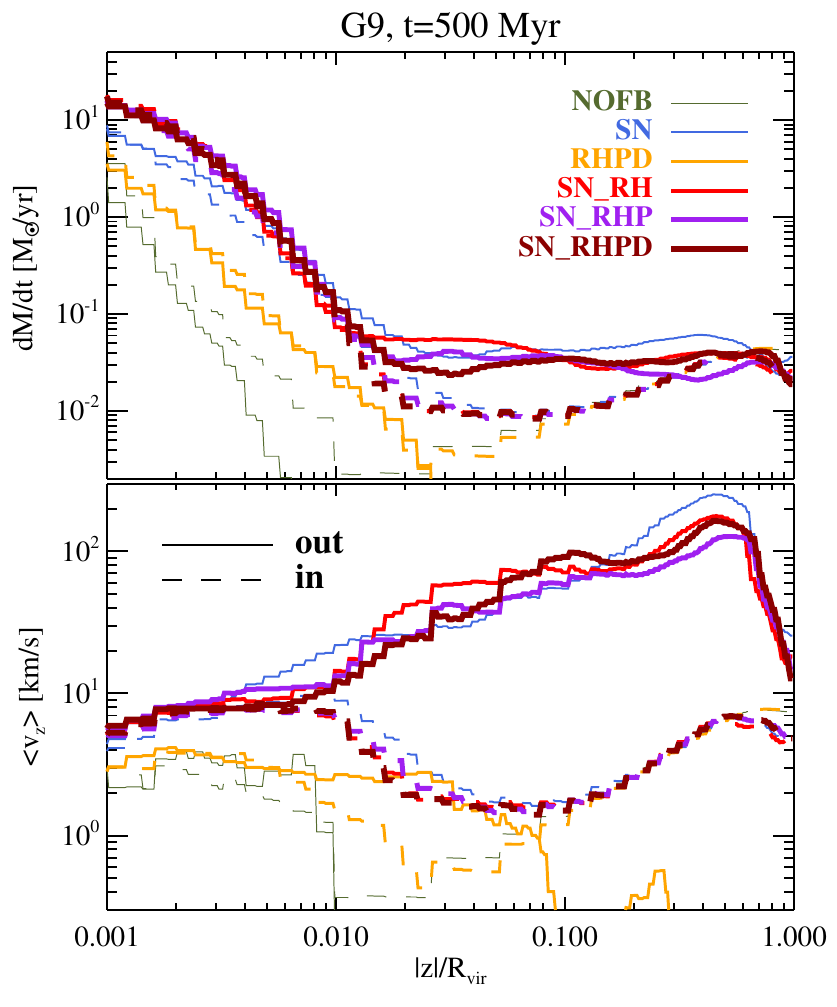}
  \caption
  {\label{OFr_G9.fig}Outflow/inflow rates (upper plot) and speeds
    (lower plot) in the \sbc{} galaxy at $500$ Myr.}
\end{figure}

In \Fig{OFr_G9.fig} we focus on the end-time of $500$ Myr and show gas
flow rates and mean speeds across parallel planes as a function of
distance $|z|$ from the galaxy plane. Here we split the gas cells into
those with outflowing and inflowing $z$-velocities, shown in solid and
dashed curves, respectively. RT feedback has very little effect on
outflow/inflow rates or speeds when added to SN feedback, except at
$|z|\la 3 \times 10^{-3} \, \Rvir \approx 300$ pc, which is more or
less inside the gas disk. At these small distances from the central
plane of the disk, RT feedback slightly increases the outflow rates,
but notably also the inflow rates, which follow the outflow rates
closely. This shows that the RT feedback has the main effect of
stirring up the disk gas without ejecting it from the galaxy. This
matches with the qualitative differences in the edge-on maps in
\Fig{maps_G9.fig}, where the SN+RT feedback runs can be seen to have a
slightly thicker and more diffuse disk than the SN only case.  By
itself, radiation feedback does not produce outflows (yellow curves in
\Fig{OFr_G9.fig}), but it thickens the disk considerably compared to
the no feedback case (green curves).

\subsubsection{The effect of the radiation} \label{rad_effect.sec} 
We found in the previous subsection that radiation feedback helps
regulate star formation in the \sbc{} galaxy. Photoionisation heating
dominates the radiation feedback, while radiation pressure appears to
have very little effect, if any.

\begin{figure*}
  \centering
  \subfloat
  {\includegraphics[width=0.47\textwidth]
    {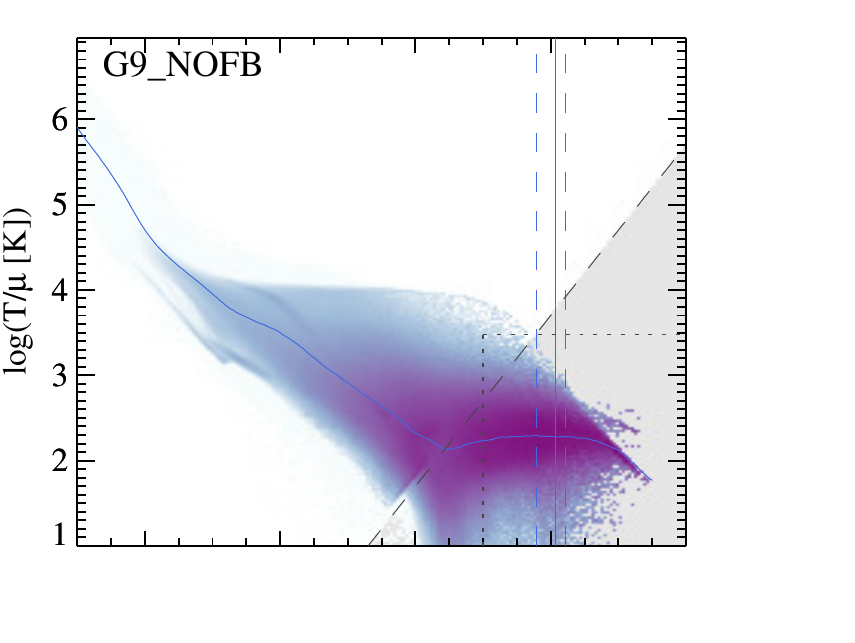}}
  \hspace{-2.4cm}
  \subfloat
  {\includegraphics[width=0.47\textwidth]
    {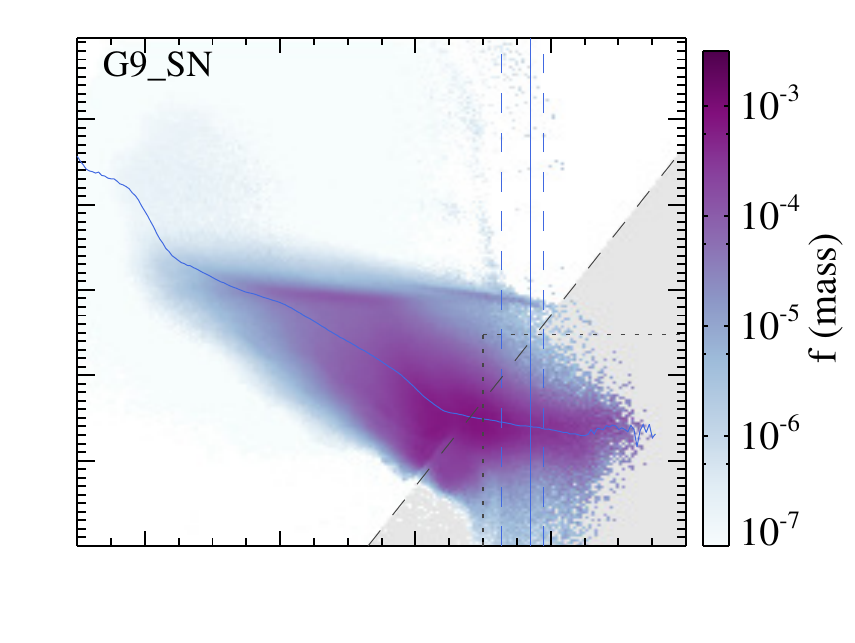}}
  \vspace{-1.25cm}
  \subfloat
  {\includegraphics[width=0.47\textwidth]
    {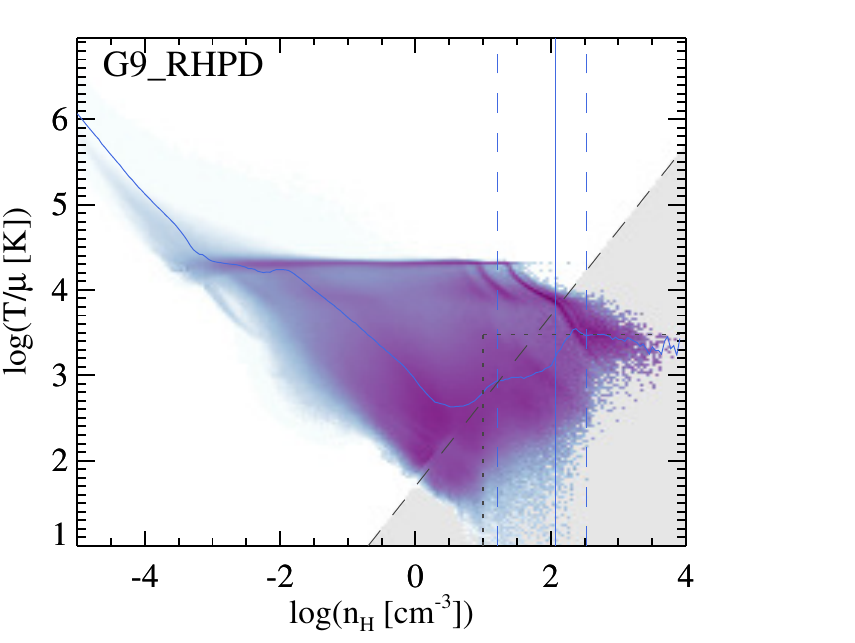}}
  \hspace{-2.4cm}
  \subfloat
  {\includegraphics[width=0.47\textwidth]
    {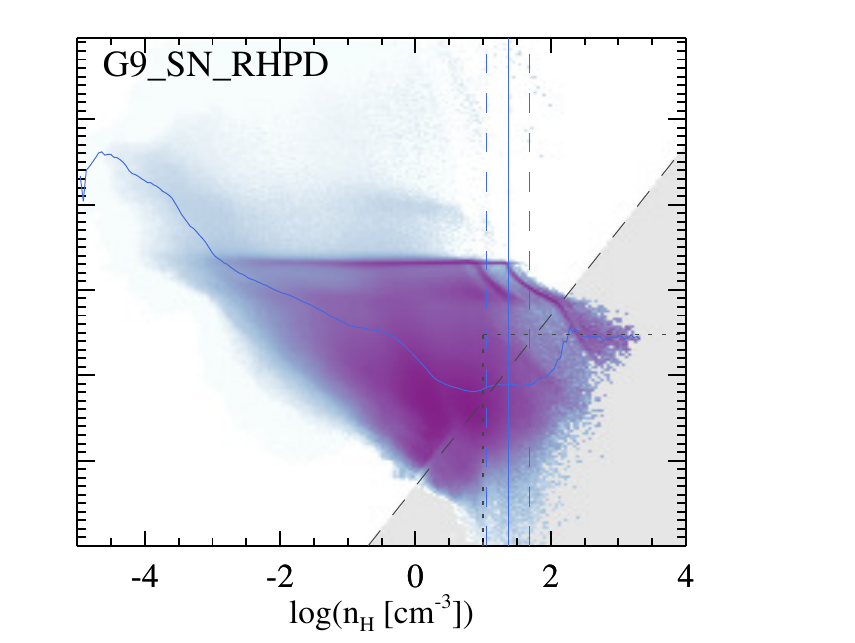}}
  \caption
  {\label{Ph_G9.fig}Temperature-density phase diagrams, time-stacked
    from snapshots every $50$ Myr after relaxation, in the \sbc{} runs
    with different feedback processes, as indicated in the top right
    corner of each plot. The vertical solid lines show the
    mass-weighted mean density in each snapshot and the solid curves
    show the median temperature in each density bin. The dotted lines
    show the temperature threshold (horizontal) and density threshold
    (vertical) for star formation. The diagonal dashed lines indicate
    the non-thermal Jeans pressure, \Eeq{polytrope.eq}, which is added
    to the other pressure terms (thermal pressure and trapped
    radiation pressure) in the hydrodynamics to prevent artificial
    fragmentation. Assuming a negligible contribution from trapped IR
    radiation, the pressure of gas below this line, as indicated by
    the shaded background color, is dominated by the artificial Jeans
    pressure, since the Jeans temperature is larger than $T/\mu$. }
\end{figure*}

We now consider how the photons affect the properties of the galactic
gas. We compare in \Fig{Ph_G9.fig} temperature-density phase diagrams
of gas in the \sbc{} galaxy, for the cases of no feedback (top left),
SN feedback only (top right), RT feedback only (bottom left), and
combined SN+RT feedback (bottom right). For the RT feedback we have
included all radiation feedback processes, but we note that removing
radiation pressure, direct or on dust, has no discernible impact on
the diagrams. 

The diagrams show stacked results from outputs every $50$ Myr for
$t=100-500$ Myr\footnote{i.e. from outputs at
  $t=100,150,200,250,300,350,400,450,500$ Myr}, starting after the
initial relaxation of star formation seen in \Fig{sfr_G9.fig}. We
stack the results to show a crude time-average and reduce the
stochastic influence of the formation and destruction of dense clouds,
which can shift the maximum densities considerably. Apart from this
shift in the maximum density tail, there is no qualitative change in
the diagrams between the stacked snapshots. We will refer to results
stacked by the same snapshots as ``time-stacked'' in the remainder of
this paper.

We over-plot the star formation thresholds in density and temperature
(vertical and horizontal dotted lines), the median temperature per
density bin (solid blue curve), and the mass-weighted mean density
(solid blue vertical line). We bracket the mean densities by the
maximum and minimum means per stacked snapshot (dashed blue vertical
lines), indicating the shift caused by the formation and destruction
of dense clouds. The diagonal dashed lines indicate the non-thermal
Jeans pressure, \Eeq{polytrope.eq}, used to prevent resolution-induced
fragmentation of gas \citep{Truelove:1997bj}. The artificial pressure
term dominates the pressure of gas below this line, i.e. in the shaded
region, making it the de-facto dominant ``feedback'' in this
high-density low-temperature gas. Without feedback (top left), the
Jeans pressure predominantly supports this gas, while adding SN
feedback (top right), RT feedback (bottom left), or a combination of
the two (bottom right), typically increases the temperature and
decreases the density, and thus reduces the amount of gas supported by
this artificial pressure. One should keep in mind throughout that the
effect of adding SN and RT feedback is somewhat weakened by the
existence of this Jeans pressure, which must be in place in all
simulations as a last resort to keep gas from collapsing beyond the
resolution limits.

In the bottom left diagram, we see that radiation feedback on its own
increases the median temperature of dense gas compared to no feedback
(top left), by heating a considerable amount of photo-ionised gas to
$\sim 10^4$ K. However, it has only a tiny impact on the mean density,
compared to the no feedback case. Combined with SN feedback, radiation
(bottom right) is much more efficient at decreasing the mean density,
by almost half a dex compared to SN feedback only. 

Judging from the diagrams, the suppression in star formation due to
radiation feedback appears to owe to either of two effects, or both:
i) direct heating of the gas, which raises it above the temperature
threshold of $3000$ K for star formation, i.e. gas moves up, or ii)
resistance to gas collapse, indirectly due to the heating, i.e. gas
moves to the left. To investigate the direct effect, we have repeated
runs \sbc\_\simf{} and \sbc\_\sime{}, after removing the temperature
threshold for star formation. The run with radiation feedback only,
i.e. \sbc\_\simf{}, shows slight sensitivity to the temperature
threshold, with $10\%$ more stellar mass formed at $500$ Myr with the
threshold removed, while the run with SN+RT feedback (\sbc\_\sime{})
actually produces $10\%$ \emph{less} stars if the temperature
threshold is removed, which owes to an increase in the SN feedback
efficiency. We conclude that the effect of radiation feedback is
primarily due to adiabatic resistance to gas collapse, rather than the
precise temperature threshold for star formation.

\begin{figure}
  \centering
  \includegraphics[width=0.45\textwidth]
    {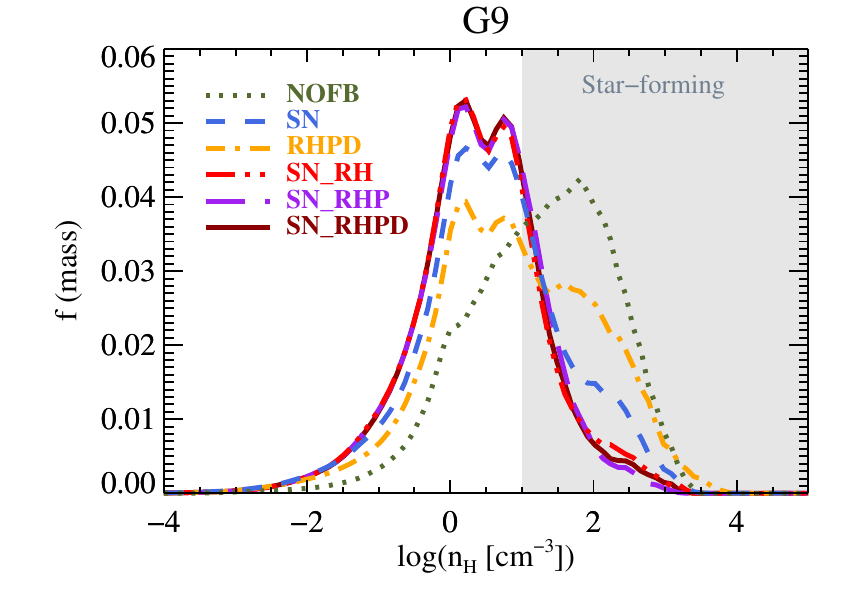}
  \caption
  {\label{hist_nH_G9.fig}Time-stacked mass-weighted density
    distribution in the \sbc{} galaxy. Star forming gas is indicated
    by the shaded region. SN and radiation feedback suppresses high
    gas densities. The suppression from radiation is dominated by
    radiation heating, but IR and optical pressure on dust provides
    marginal extra support.}
\end{figure}

The bottom phase diagrams of \Fig{Ph_G9.fig} reveal conspicuous
features at the right end of the photo-ionised temperature plateau
($\approx 2 \times 10^4$ K), where the gas temperature decreases and
density increases along narrow tracks. They are due to the HII regions
being unresolved. The highest temperature tracks consist of single
cells filled with radiation at a constant luminosity of a single young
stellar particle, and can be accurately reproduced in single cell
tests. The lower temperature tracks consist of cells adjacent to those
source cells into which the constant luminosity propagates, again at
roughly a constant rate. Under-resolved \hii{} regions are also
visible in \Fig{G9_photons.fig}, indicated by a high contrast and
``pixelated'' peaks for the ionising photon groups in the three
left-most panels. We will return to this resolution issue in
\Sec{HIIreg.sec}.

\Fig{hist_nH_G9.fig} shows the time-stacked mass-weighted density
distribution of the gas in the \sbc{} runs. Radiation and SN feedback
clearly reduces the maximum gas density, but radiation pressure, when
added, has very little effect. The plot supports the previous
conclusion that the effect of the radiation heating is to prevent
collapse of the gas by increased thermal pressure, which keeps the gas
at lower densities. The effect is quite mild though, as the change in
the density distribution is small when radiation is added to SN
feedback.

\begin{figure}
  \centering
  \includegraphics
    {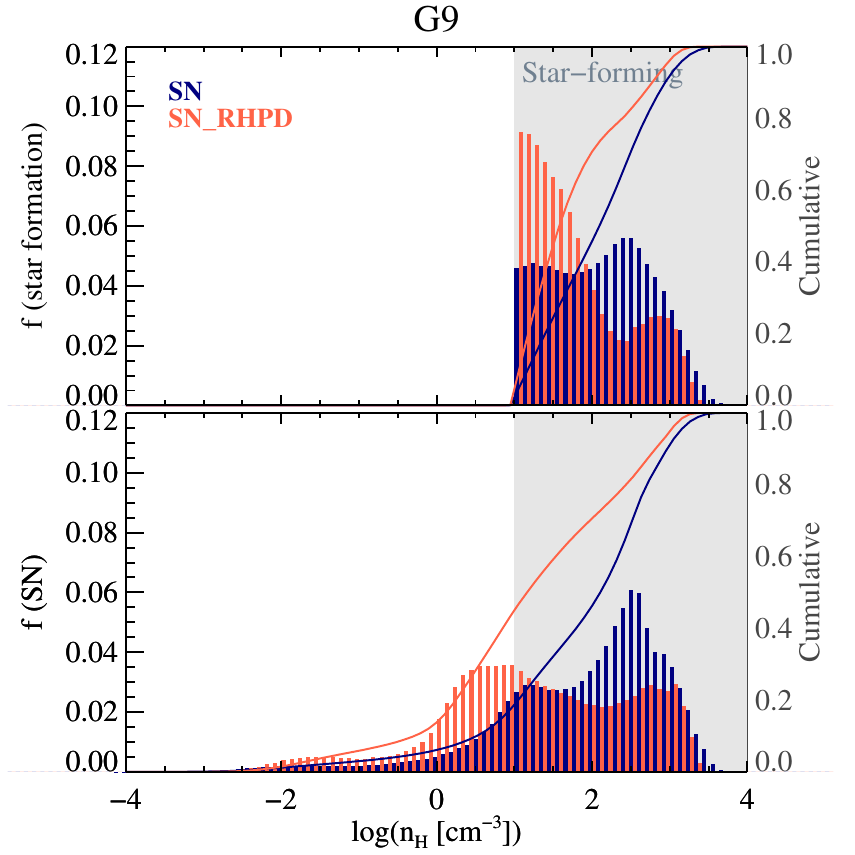}
  \caption
  {\label{star_rhohist.fig}Comparison between the SN and SN+RT \sbc{}
    runs, of the probability distribution functions for the gas
    density at which stellar particles are created (upper panel) and
    produce SNe (lower panel). The shaded regions indicate star-forming
    densities. The solid curves in each panel show the cumulative
    probabilities. The upper panel indicates that RT feedback lowers
    the densities at which stellar particles are born, which should
    increase the SN feedback efficiency by allowing SN events to take
    place in a lower-density medium, as verified in the bottom panel.}
\end{figure}

\begin{figure}
  \centering
  \includegraphics
    {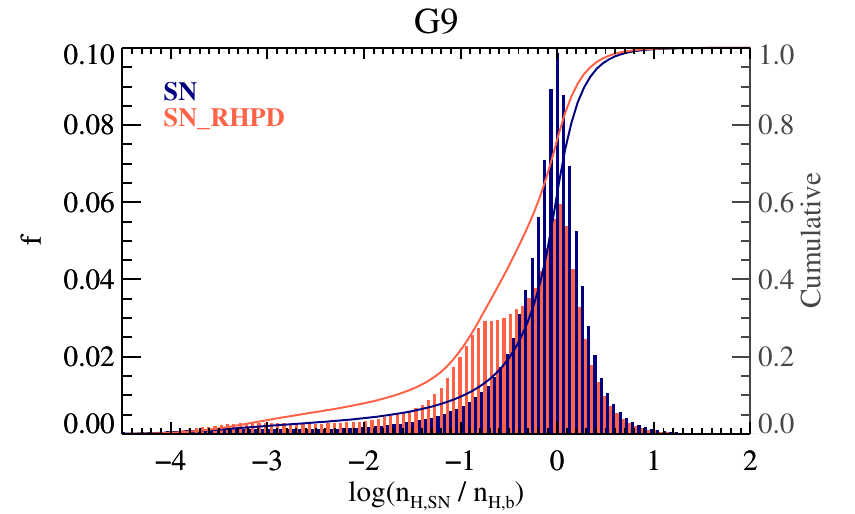}
  \caption
  {\label{star_drhohist.fig}Comparison between the \sbc{} runs with SN
    and SN+RT, of the probability distribution functions for the
    increase/decrease in the surrounding gas density between stellar
    particle birth and SN event. A value of zero at the x-axis
    indicates that the surrounding gas density stays unchanged from
    birth to SN, while negative/positive values correspond to a
    decrease/increase in density. The solid curves show the cumulative
    probabilities. RT feedback has the effect of somewhat, but not
    dramatically, diffusing the gas around the stellar particles,
    before they produce SNe.}
\end{figure}

\subsubsection{SN amplification}\label{SN_boost.sec}
Radiation can plausibly have the effect of amplifying SN feedback
\citep{Pawlik:2009jv, Geen:2015bs}. It can diffuse the surrounding
gas, which has the well-known effect of decreasing the cooling rate,
which scales with density squared. This in turn can make SNe more
effective in stirring up the ISM, suppressing star formation and
generating outflows. This may happen as a combination of two effects:
by preconditioning of the medium by the radiation before the SN events
take place, but also in a preventive way, where the radiation feedback
shifts the typical star formation densities to lower values, which
directly causes SN events to take place at lower densities. 

\Fig{star_rhohist.fig} shows the probability distribution of gas
densities at which stellar particles are formed (upper panel) and at
which they produce SNe $5$ Myr later (lower panel), in the \sbc{} runs
with SN only and with full RT feedback added. From the upper panel we
see that the RT feedback shifts star formation to lower densities,
which now peak at the star formation threshold, whereas they peak
$1.5$ dex above the threshold with SN feedback only. One also can read
from the cumulative probability curves (solid lines) that with SN
feedback only, about $45\%$ of the stars are formed at
$\nh \la 10^2 \, \cci$ (ten times the star formation threshold,
$\ns$), while $\approx 70\%$ of the stars form below the same density
with RT feedback added. This then translates into a similar difference
in the SN densities in the lower plot. With SN feedback only, about
$45\%$ of the stars produce SNe in gas with densities below
$10 \, \ns$, while the addition of RT feedback increases this to
$70\%$. This similarity in the characteristic density difference
between the two plots indicates that preconditioning of the gas by
radiation does not play a major role. If it did, we should expect the
typical SN densities to shift even further to lower densities.

Even so, we take a closer look at the effect of radiation
preconditioning in \Fig{star_drhohist.fig}. Here we plot the
probability distribution functions, for SN and SN+RT feedback, for the
relative \emph{difference} between surrounding densities at stellar
particle birth, $n_{\rm H,b}$, and SN event, $n_{\rm H,SN}$. The idea
is that we remove the effect of the stars being born at lower
densities with RT feedback. For the SN feedback only case, we find a
strong peak in the probability around $n_{\rm H,SN}/ n_{\rm H,b}=1$,
which just means that typically a stellar particle's birth and SN
event happen at the same density. A slight majority, $\approx 60\%$,
of the stars produce SNe at lower densities, and there is a tail in
the distribution with a few percent of the SNe exploding at orders of
magnitude lower densities. With radiation feedback added, the peak is
still in the same place, but the distribution and the tail is shifted
towards lower densities. The effect is not large though.

In addition to giving information about the nature of (possible) SN
amplification by radiation feedback, Figures \ref{star_rhohist.fig}
and \ref{star_drhohist.fig} give us a hint about how radiation
feedback suppresses star formation. The radiation shifts star
formation to substantially lower densities (\Fig{star_rhohist.fig}),
but does not as substantially diffuse gas locally around young stellar
particles (\Fig{star_drhohist.fig}), suggesting that the effect of
radiation feedback is more to \emph{prevent} the formation of dense
clumps, rather than destroying them after they form.

\subsection{G10: Milky Way mass galaxy}
\begin{figure}
  \centering
  \subfloat
  {\includegraphics[width=0.47\textwidth]
    {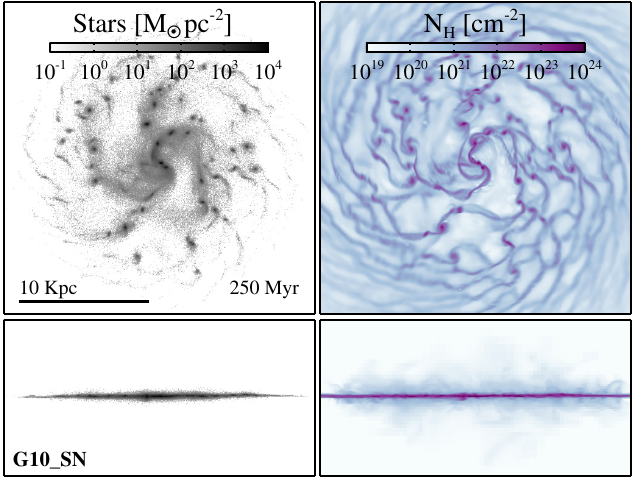}}
  \vspace{0.5mm}
  \subfloat
  {\includegraphics[width=0.47\textwidth]
    {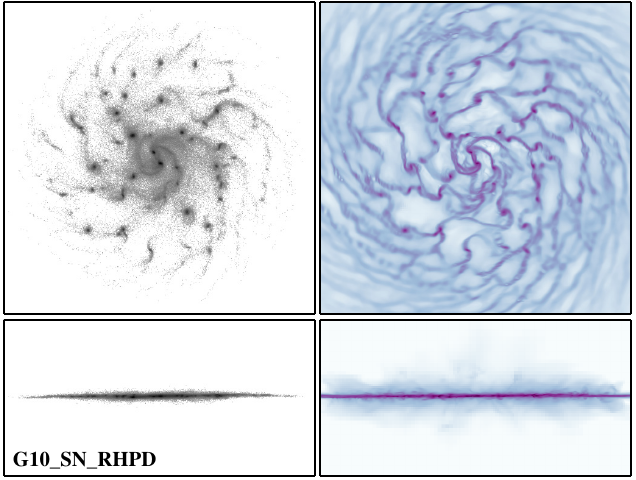}}
  \caption
  {\label{maps_G10.fig} Maps of the \mw{} galaxy (baryonic mass of
    $3.5 \times 10^{10} \ \Msun$) at $250$ Myr, for SN feedback only
    (upper panel) and full radiation feedback (lower panel). Each
    panel shows face-on and edge-on views of the stellar density
    (left) and total hydrogen column density (right). Radiation
    feedback has little noticeable effect in this galaxy, and in fact
    the same applies for SN feedback (the comparison to no feedback is
    not shown).}
\end{figure}

We now turn our attention towards our most massive galaxy, similar in
mass to the Milky Way (MW) Galaxy. The galaxy is ten times more
massive than the \sbc{} galaxy we have analysed so far, and of
interest here is how the galaxy mass affects the radiation
feedback. The mass is not the only thing different from the \sbc{}
galaxy, however. The metallicity of the gas is ten times higher and
the gas fraction is considerably less: $30\%$, compared to $50\%$ for
the \sbc{} galaxy. It makes sense to change also these properties,
since the idea is to roughly follow the stages in the evolution of the
present day MW. However, in \Sec{metmass.sec} we will disentangle the
effects of these different galaxy properties on the radiation
feedback.

\begin{figure}
  \centering
  \includegraphics
    {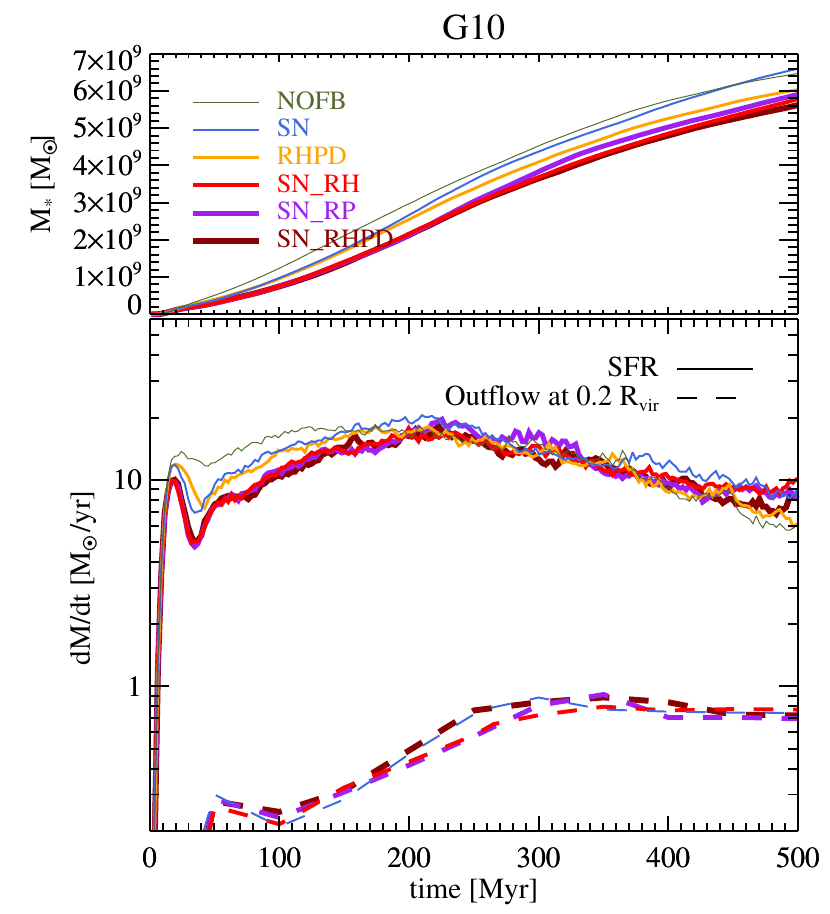}
  \caption
  {\label{sfr_G10.fig}Star formation and outflow rates in the \mw{}
    runs with different feedback processes included, as indicated in
    the legend. {\bf Upper panel:} stellar mass formed over time. {\bf
      Lower panel:} star formation rates (solid lines) and outflow
    rates across planes at distances of $0.2 \ \Rvir$ from the disk
    plane (dashed lines). Feedback is much less effective here than in
    the less massive \sbc{} galaxy (cf. \Fig{sfr_G9.fig}). Radiation
    heating suppresses star formation more than SN feedback, but the
    effect is small. Radiation pressure is unimportant. Outflows are
    not affected by the radiation feedback.}
\end{figure}

We first consider the qualitative effect of radiation feedback on the
galaxy morphology in \Fig{maps_G10.fig}, where we compare face-on and
edge-on maps at $250$ Myr. We find no visible effect from radiation
pressure, neither direct nor on dust, so we only compare here the case
with SN feedback only (\mw\_\simb{}) and SN + full RT feedback
(\mw\_\sime{}). The overall effect of adding RT feedback is less than
in the \sbc{} galaxy (\Fig{maps_G9.fig}), though the disk does become
slightly less clumpy and more diffuse compared to SN feedback only. We
do not show the no feedback case, but it looks very similar to the
case with SN feedback, so also SN feedback is weak in this massive
galaxy.

We go on to compare the star formation, in \Fig{sfr_G10.fig}. Here we
again see that all the modelled feedback processes are much weaker
than in the previous less massive galaxy. SN feedback initially
slightly reduces the star formation compared to the no feedback case,
but ends up with \emph{more} stars formed (which is due to the
recycling of gas in the SN case, resulting in an effectively larger
gas reservoir). In such a massive galaxy, SN feedback is though to
become decreasingly important, and AGN feedback, which is not
modelled, may start to dominate \citep{Bower:2006fj}. Also, it is
likely that numerical overcooling becomes stronger, due to the
increasing gravitational potential, gas densities, metallicity, and
decreasing physical resolution (although the larger stellar particle
mass should somewhat compensate by injecting more energy per SN
event).

In this galaxy, radiation feedback has a stronger effect on the star
formation than SNe, though the effect is still weak, with an
$\approx 7 \%$ reduction in the stellar mass formed (at $500$ Myr)
with RT feedback only, and $\approx 10 \%$ if combined with SN
feedback. The slightly increased feedback efficiency when combined
with SNe hints at an amplification effect, but the effect is
small. Radiation heating dominates the radiation feedback, as adding
radiation pressure and dust interactions has little effect on the star
formation. 

Outflows rates across planes $0.2 \, \Rvir$ ($38.4$ kpc) from the disk
are shown by dashed lines in the bottom panel of
\Fig{sfr_G10.fig}. Outflows appear to be powered nearly exclusively by
SN feedback, since the rates remain virtually unchanged after the
addition of radiation feedback (of any sort). The mass loading factor
of the outflow remains at $\la 0.1$, similar to the \sbc{} galaxy.

\subsection{G8: gas-rich dwarf}
\begin{figure}
  \centering
  \subfloat
  {\includegraphics[width=0.47\textwidth]
    {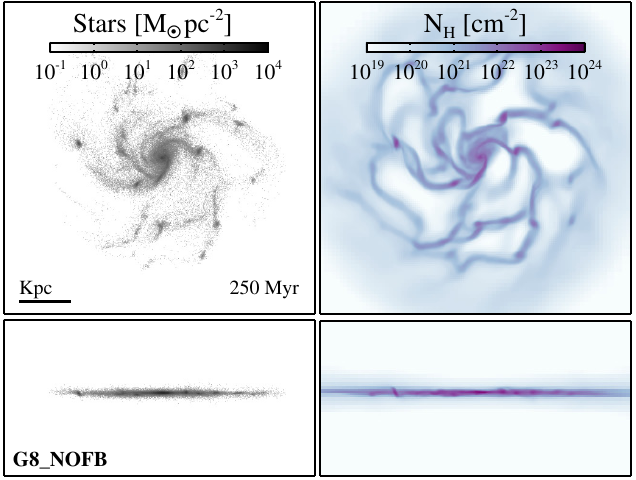}}
  \vspace{1.mm}
  \subfloat
  {\includegraphics[width=0.47\textwidth]
    {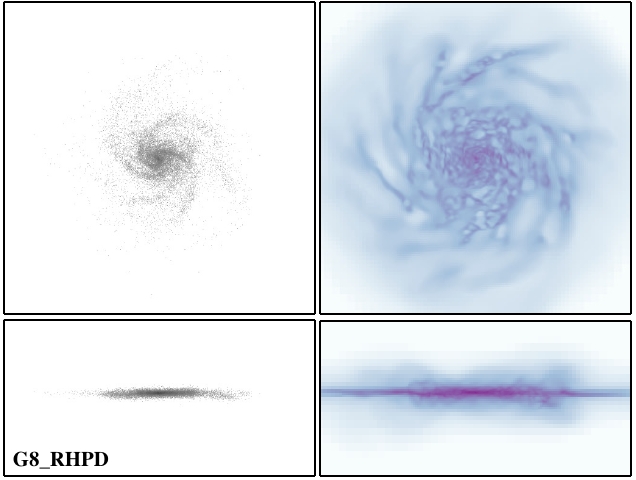}}
  \vspace{1.mm}
  \subfloat
  {\includegraphics[width=0.47\textwidth]
    {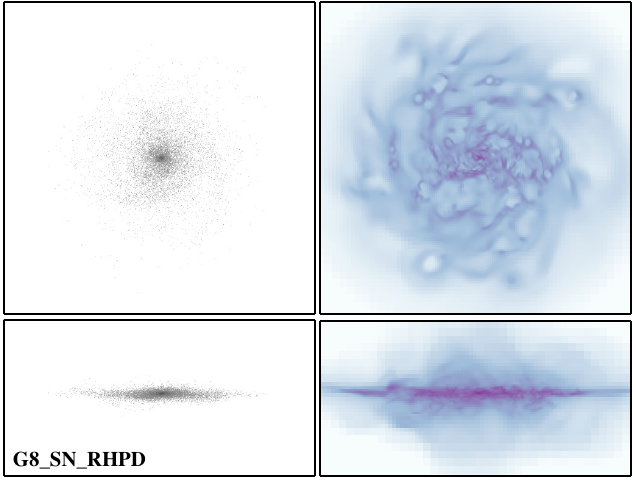}}
  \caption
  {\label{maps_G8.fig} Face-on and edge-on maps of stellar density
    (left) and total hydrogen column density (right) for the \dw{}
    galaxy (baryonic mass of $3.5 \times 10^{8} \ \Msun$) at $250$
    Myr, with no feedback (top panel), full radiation feedback (middle
    panel), and added SN feedback (bottom panel). Radiation feedback
    alone efficiently prevents the formation of massive clumps. SN
    feedback alone (not shown) has a similar qualitative effect,
    though it results in a slightly thicker gas disk. Combining the RT
    and SN feedback smooths the gas distribution further.}
\end{figure}
\begin{figure}
  \centering
  \includegraphics
    {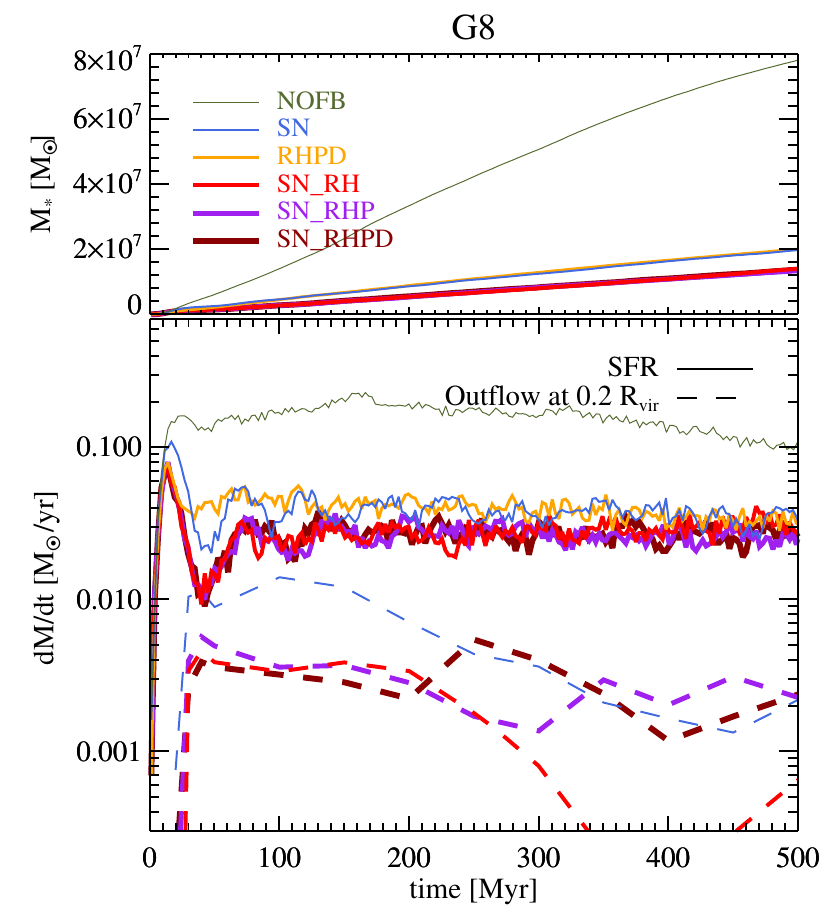}
  \caption
  {\label{sfr_G8.fig}Star formation and outflow rates in the \dw{}
    runs with different feedback processes included, as indicated in
    the legend.  {\bf Upper panel:} stellar mass formed over
    time. {\bf Lower panel:} star formation rates (solid lines) and
    outflow rates across planes at distances of $0.2 \ \Rvir$ from the
    disk plane (dashed lines). Radiation feedback is as effective as
    SN feedback at reducing star formation, but does it more smoothly,
    with SN feedback resulting in more bursty star formation. The
    relative effect of including a single feedback process (RT or SN)
    is much stronger than that of adding a second one. Radiation
    heating dominates the suppression in star formation, and reduces
    both the outflow rate and the mass loading factor of the outflow.
    Radiation pressure has a negligible effect on the star formation,
    but increases the outflow rate during the final $\approx 250$
    Myr.}
\end{figure}

We now consider variations with RT feedback in the least massive
galaxy, \dw{}. Its properties only differ from those of the
intermediate mass \sbc{} galaxy in terms of the halo and galaxy mass.
The gas fraction and metallicity are unchanged, at $50\%$ and
$0.1 \, \Zsun$, respectively.

We begin with a qualitative comparison of morphologies with the
inclusion of different feedback processes, shown in
\Fig{maps_G8.fig}. We compare the cases of no feedback (top panel)
full RT feedback (i.e. heating, direct and dust pressure, middle
panel), and SN+RT feedback (bottom panel). RT feedback on its own is
clearly much more efficient in this galaxy than in the previous, more
massive ones. It completely suppresses the formation of massive
clumps, smooths out density contrasts, and considerably reduces the
formation of stars. We do not show the case with SN feedback only, but
note that in the weak gravitational potential of the \dw{} galaxy, it
has a similar qualitative effect as RT feedback only, with the only
clear difference being a somewhat thicker gas disk for SN
only. Combining RT and SN feedback, however, has some additional
impact on the galaxy morphology, with fewer stars and thicker, more
diffuse gas disk (bottom panel of \Fig{maps_G8.fig}).

We compare the star formation rates and outflows for the \dw{} galaxy
in \Fig{sfr_G8.fig}. Here we see that the star formation rates with RT
feedback only are very similar to those in the SN only case. The
combination of SN and RT feedback reduces the star formation by about
$25 \%$ compared to including only one of those processes, which is
much less than the relative reduction from the no feedback case when
either process was added, which is $\approx 75 \%$.  In
\Sec{rad_effect.sec} we searched qualitatively for the existence of a
feedback amplification in the \sbc{} galaxy, i.e. where the addition
of one form of feedback (RT) boosts the efficiency of another form
(SNe) in quenching star formation, but found no clear evidence. Here
we have an indication of the opposite effect.

The inclusion of direct radiation pressure and dust interactions has
no effect on the star formation rate. However, unlike the case of the
more massive galaxies, it increases the mass outflow rates
non-negligibly, restoring the outflow rate at late times back to that
obtained with SN feedback only, as shown by dashed lines in the lower
panel of \Fig{sfr_G8.fig}. The effect comes predominantly from direct
pressure from the ionising radiation, as can be seen by comparing the
purple and dark red dashed curves.

\subsection{All galaxies: metallicity versus mass}\label{metmass.sec}
Comparison of the three galaxies in the previous subsections reveals a
clear trend: the efficiency of RT feedback decreases with increasing
galaxy mass. However, we varied not only the mass, but used a ten
times higher metallicity in the \mw{} galaxy than in the less massive
ones. We now want to investigate how much the RT feedback efficiency
is affected by galaxy mass, i.e. the gravitational potential, and how
much by the gas metallicity, via its influence on the gas cooling
time. For this purpose, we have run the three galaxies at \emph{both}
the metallicities we have considered, i.e. $1 \, \Zsun$ and
$0.1 \, \Zsun$.

\begin{figure}
  \centering
  \includegraphics[width=0.48\textwidth]
    {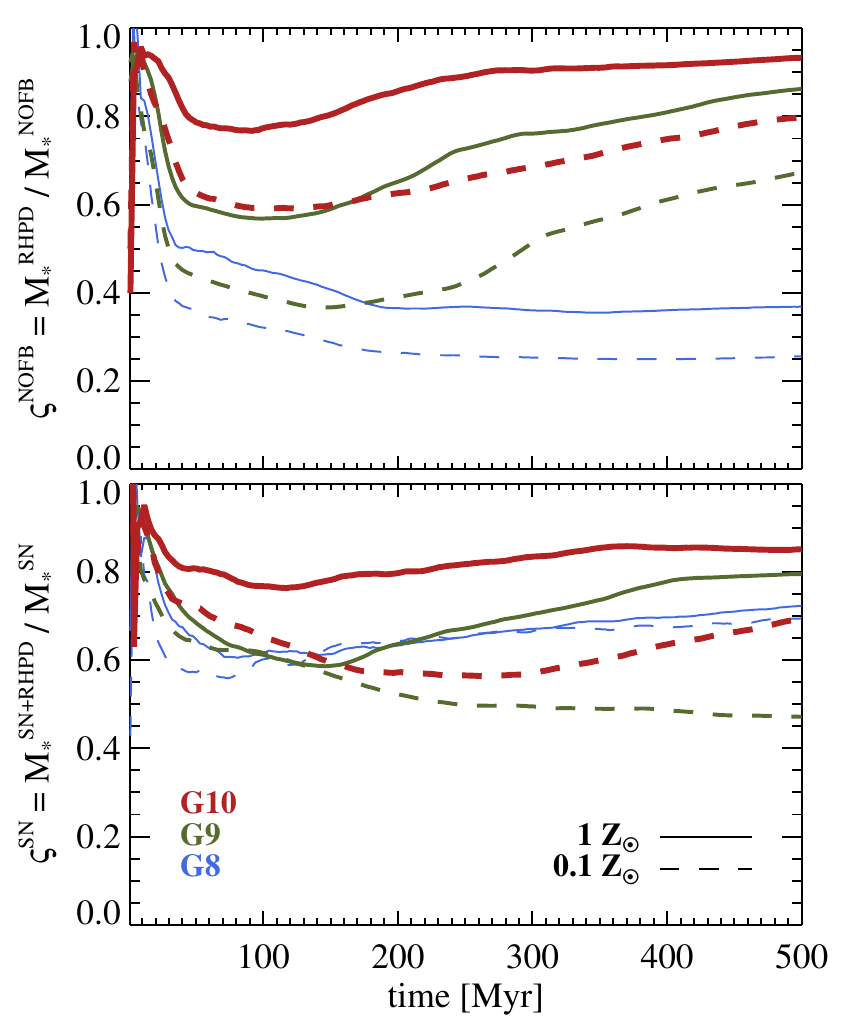}
  \caption
  {\label{FBeff.fig}RT feedback efficiency, i.e. cumulative
    suppression of star formation due to RT feedback (\simf), plotted
    against time and compared for different galaxy masses and
    metallicities, as indicated in the legend. Different metallicities
    are denoted by solid and dashed curves, while galaxy mass is
    denoted by color and thickness, with increasing thickness
    indicating higher mass. An efficiency value of $1$ corresponds to
    no effect on the star formation, while a value close to $0$
    indicates a strong reduction of the star formation. {\bf Upper
      panel:} RT feedback efficiency when acting alone. {\bf Lower
      panel:} RT feedback efficiency when combined with SN
    feedback. Both increased mass and metallicity reduce the
    efficiency of radiation feedback (and SN feedback, not shown),
    except if radiation is combined with SN feedback, where the
    efficiency peaks for the intermediate galaxy mass.}
\end{figure}

We quantify the efficiency of radiation feedback by calculating the
relative reduction of stellar mass formed when a reference simulation
is re-run with the addition of full RT feedback, i.e.
\begin{align}
  \zeta^{\rm X}(t) = \frac{\Mstar(t)^{\rm X+RHPD}}{\Mstar(t)^{\rm X}},
\end{align} 
where $\Mstar(t)$ is the stellar mass formed in the simulation up to
time $t$ and X represents the feedback included in the reference
simulation. Values of $\zeta^{\rm X}<1$ correspond to feedback which
suppresses star formation, while $\zeta^{\rm X}>1$ indicates
\emph{negative} feedback, i.e. enhanced star formation. In
\Fig{FBeff.fig} we plot two such RT feedback efficiencies,
$\zeta^{\rm NOFB}$ in the upper panel, which shows the factor by which
RT feedback suppresses the stellar mass relative to the simulation
without feedback, and $\zeta^{\rm SN}$, in the lower panel, which
shows the suppression when RT feedback is added to SN feedback.

The upper panel shows the effect of radiation feedback in isolation,
and gives a ``cleaner'' indication of the feedback efficiency than the
lower panel, where the curves are quite sensitive to SN feedback
efficiency, which is also (and independently) sensitive to the galaxy
mass, metallicity, and stellar particle mass\footnote{Another factor,
  which we have not considered so far, is the effect that the stellar
  particle mass has on RT feedback. We have investigated this for one
  of our galaxies, as discussed in \App{restests.sec}. The indication
  there is that while stellar particle masses have a large effect on
  the SN feedback efficiency, they have much less impact on the RT
  feedback, which is likely because the energy injection is smooth
  rather than instantaneous.}. However, the lower panel is quite
important, since the \emph{addition} of radiation to SN feedback is
more physically relevant than considering radiation feedback in
isolation.  We see from both panels that \emph{both} increasing galaxy
mass and metallicity weaken the effect of RT feedback (the same
applies for SN feedback, though this is not shown in these plots).

The emerging qualitative picture is as follows: star formation in low
mass galaxies is easily regulated by SN feedback, due to a combination
of long cooling times (low metallicity and density), weak
gravitational potential, and relatively massive stellar
particles. Although RT feedback is by itself roughly as effective at
regulating star formation (see \Fig{sfr_G8.fig}), adding it to SN
feedback has relatively little effect on the SF regulation, reducing
the star formation rate by a few tens of percent (\Fig{FBeff.fig},
lower panel)\footnote{\cite{Hopkins:2012ez} find qualitatively similar
  non-linear effects when combining feedback processes.}. With
increasing galaxy mass, both SN and RT feedback become less effective
(\Fig{sfr_G9.fig} and upper panel of \Fig{FBeff.fig}), but combining
them may have a larger effect (\Fig{FBeff.fig}, lower panel), though
this is quite sensitive to the metallicity. At even higher mass, the
gravity and cooling has become strong enough that not even the
combination of feedback processes can significantly halt the star
formation (Figs. \ref{sfr_G10.fig} and \ref{FBeff.fig}), especially at
the higher, and more realistic, metallicities.

\subsection{IR multi-scattering}\label{mscatter.sec}
\begin{figure}
  \centering
  \includegraphics
    {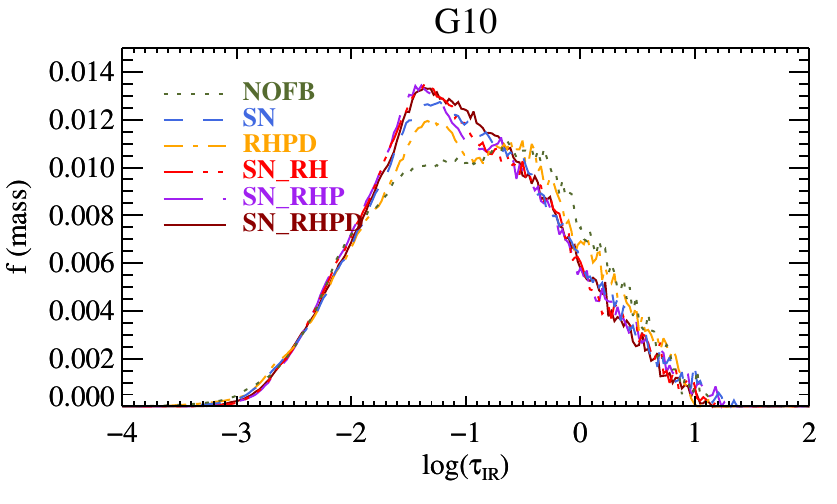}
  \caption
  {\label{distTau_G10.fig}Time-stacked mass-weighted probability
    distribution of IR optical depth, $\tauIR$ along LOSs through the
    face of the \mw{} galaxy. The different curves represent inclusion
    of different feedback processes, as indicated in the legend. A
    fraction of the gas in this galaxy is in the optically thick
    regime, $\tauIR>1$, where multi-scattering starts to play a
    role. However, it has little impact on the galaxy, as can be
    deduced from the similarity of the optical depth distributions
    with IR multi-scattering pressure excluded and included (purple
    dot-long-dashed and solid dark red, respectively).}
\end{figure}
The pressure due to multi-scattering IR photons has been cited as an
important radiation feedback process \citep{Thompson:2005dq,
  Murray:2010gh, Hopkins:2012ez, Hopkins:2012hr, Agertz:2013il}, yet
we do not appear do get much of an effect at all from radiation
pressure, including that of the IR radiation. In \Fig{distTau_G10.fig}
we show the mass- (or column density-) weighted distributions of LOS
IR optical depths, $\tauIR$, through the face of the \mw{} galaxy,
which has the largest optical depths of our disks. The LOSs are taken
from time-stacked snapshots (every $50$ Myr starting at $100$ Myr, as
usual), and each has the width of the finest AMR resolution, or $36$
pc. The plot quantifies how the mass is distributed in face-on optical
depths, which can safely be assumed to be consistently lower than
edge-on optical depths, and thus more relevant for estimating the
number of scatterings ($\approx \tauIR$) photons typically experience
before escaping the disk.

A non-negligible fraction of the gas mass has larger than unity IR
optical depths, so radiation trapping and multi-scattering does take
place in the \mw{} galaxy, with \emph{maximum} values of
$\tauIR \approx 10$.  However, as we can clearly see by comparing the
curves in \Fig{distTau_G10.fig}, these opacities are not large enough
for the IR radiation to diffuse the gas (and hence decrease the
optical depths).  Due to the lack of resolution, the gas does not
reach the high densities, and hence optical depths, where
multi-scattering plays a significant role. We can contrast these
results to those of \cite{Hopkins:2011fk}, where typical optical
depths around young stellar particles are found to be much higher,
$\sim 10-100$, in HD simulations with $\sim$pc resolution.

\cite{Murray:2011en} argued that the \emph{collective} radiation
pressure from star formation can generate cold ($\approx 10^4$ K)
outflows. Although our resolution is insufficient to resolve each
individual optically thick cloud, we should in principle see this
collective large-scale effect in our simulations, but we do not. The
\cite{Murray:2011en} argument applies to massive starbursting
galaxies, and a critical star formation rate surface density of
$\dot{\Sigma}^{\rm crit}_{*} \approx 0.1 \ \Msun \ {\rm yr}^{-1} \
{\rm kpc}^{-2}$.
Our galaxies do reach
$\dot{\Sigma}_{*} \sim 10 \ \dot{\Sigma}^{\rm crit}$, but this is
confined to clumps and centers, with most of the disk below the
critical value. However, even if the star formation is mostly below
the critical value, we would expect to see some effect of the
radiation on outflows, and it is thus interesting that we see no clear
effect at all.

Resolution may still be the defining issue though:
\cite{Murray:2011en} envision neutral clouds where the radiative force
acts on the surface facing the disc. We do not resolve these dense
clouds, and radiation momentum is deposited more smoothly throughout
whatever neutral gas exists in the halo. In \Sec{whatif.sec} we
explore qualitatively what we can expect with better resolution, by
artificially increasing the IR opacity (but find that outflows are
still not generated).

\section{Discussion} \label{Discussions.sec} Summarising the results,
we find that radiation feedback has a modest effect on the star
formation rates of our simulated galaxies, while outflows are more or
less unaffected. The suppression of star formation is due to the
suppression of the formation of dense clumps. Radiation feedback
becomes less efficient with higher galaxy mass or metallicity, while
the combination of radiation and SN feedback appears most effective at
intermediate masses (and low densities). Photoionisation heating
dominates the effect from radiation feedback, while radiation
pressure, whether direct or from reprocessed, multi-scattering, IR
radiation, has a negligible effect.

We will now discuss several aspects of our findings, starting with the
apparent inability to resolve \hii{} regions, as implied by
\Fig{Ph_G9.fig}. We will then validate our results analytically,
comparing the relative impact of the different radiation feedback
processes, as expected in the numerical framework. Next, we will
consider the expected effect of efficient IR feedback on star
formation and outflows by artificially increasing the IR
opacity. Finally, we will qualitatively compare our results to
previous work on radiation feedback on galaxy scales, where the
radiation effect is usually (but not always) modelled with subgrid
recipes in pure HD simulations.

\subsection{On unresolvable {\hii{}} regions}\label{HIIreg.sec}
Our simulations show indications that \hii{} regions are not resolved
at gas densities $\nh \ga 10 \, \cci$ (see \Fig{Ph_G9.fig} and
\Sec{rad_effect.sec}), which potentially affects our results at these
high densities. We consider here in detail at what limit \hii{} region
resolution becomes an issue.

\begin{figure}
  \centering
  \includegraphics
    {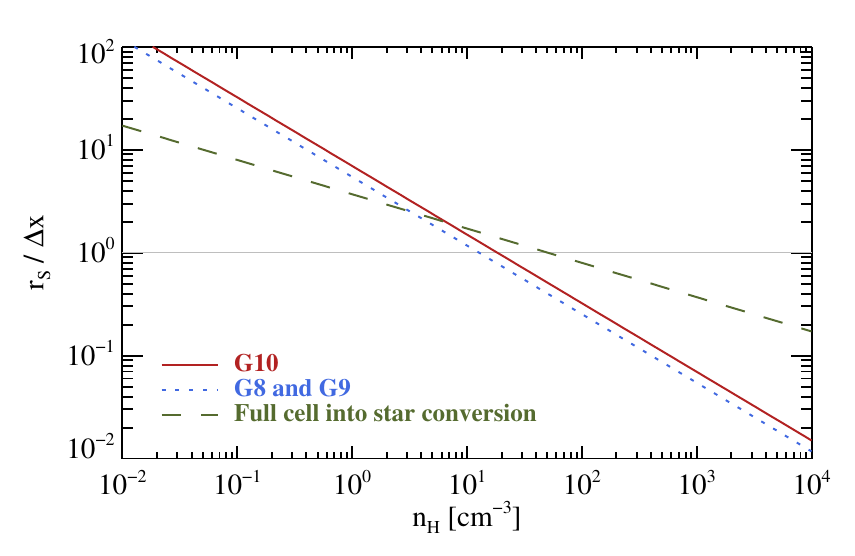}
  \caption
  {\label{rs.fig}The plot shows the ratio of the \stromgren{} radius,
    $\rS{}$, of ionised regions versus the maximum cell resolution,
    $\dx=18$ pc in \dw{}/\sbc{} (dotted blue) and $\dx=36$ pc in \mw{}
    (solid red). The \hii{} regions are not resolved above the star
    formation density threshold ($\nh > 10 \ \cci$) in our
    simulations.  The dashed green curve shows the ratio, at location
    of birth, if the total gas mass of a cell is always converted
    directly into a stellar particle (ignoring the mass depletion of
    the cell). It demonstrates that with the current star formation
    method, it is impossible to resolve \hii{} regions above the
    density threshold for star formation, if this treshold is
    $\ga 50 \ \cci$.}
\end{figure}

We can investigate this using the analytic expression for the
\stromgren{} radius of a photo-ionised region in a uniform medium. The
specific ionising luminosity of stellar sources is
$\LumNumSpec_{\rm UV}\approx5 \times 10^{46}$ ionising photons per second
per Solar mass, according to the \cite{Bruzual:2003ck} model and
assuming the \cite{Chabrier:2003kia} IMF (see \Fig{groups.fig}). A
stellar particle of mass $\mstar$ then has luminosity
$\LumNum_{\rm UV}=\LumNumSpec_{\rm UV}\mstar$ (photons per
second). The \stromgren{} radius around a stellar source is
\citep{Stromgren:1939dq}
\begin{align} \label{rs.eq}
  \rS &= \left( 
    \frac{3 \LumNum_{\rm UV}}{4 \pi \alpha^{\rm{B}} \nh^2}
  \right)^{1/3} \\
  &= 21 \ {\rm{pc}} \nonumber \\ 
  &\times \ \left( \frac{\LumNumSpec_{\rm UV}}{5 \times 10^{46} \ \sm
    \ \Msun^{-1}} 
  \frac{\mstar}{600 \ \Msun } \right)^{1/3}
  \left( \frac{\nh}{10 \ \cci} \right)^{-2/3}, \nonumber
\end{align}
where $\alpha^{\rm{B}}= 2.6 \times 10^{-13} \ \cc \ \sm$ is the case B
recombination rate of hydrogen around $10^4$ K \citep{Ferland:1992bu},
and where we have substituted the stellar particle mass used in the
\dw{} and \sbc{} simulations, along with the star formation density
threshold used in all our simulations. \Fig{rs.fig} shows the ratio of
the \stromgren{} radius and the cell width for all three simulated
galaxies. For all simulations, the \hii{} regions around young stars
are only resolved in gas below the star formation density, $n_{*}$,
and the stars must travel to densities of $\nh \la 1 \, \cci$ within
their ``luminous'' lifetime of $\approx 5$ Myr to have their \hii{}
regions resolved by more than ten cell widths.

Judging from \Eeq{rs.eq}, a simple solution to forcing \hii{} regions
to be resolved in AMR simulations would be to increase the mass (and
hence luminosity) of the stellar particles. However, unless changes
are made to the star formation recipe, we hit a concrete upper limit
in $\mstarm$, which is the total gas mass of the hosting cell,
\begin{align} \label{mmax.eq}
  \mstarm = \frac{\nh \mp}{X} \left( \dx \right)^3 = 
  1875 \ \Msun  \frac{\nh}{10 \, \cci} 
  \left( \frac{\dx}{18 {\rm pc}} \right)^3,
\end{align}
where $X=0.76$ is the hydrogen mass fraction.  Assuming, for arguments
sake, that we always convert the total gas mass of cells into stars,
and that their neighbourhood has a roughly homogeneous gas density,
the ratio of the \stromgren{} radius and cell width around newly
formed stellar particles would be (from Eqs. \ref{rs.eq} and
\ref{mmax.eq}):
\begin{align}\label{rsdx.eq}
  \frac{\rS}{\dx} &= 1.0 \ \left( 
    \frac{\nh}{50 \ \cci}  \right)^{-1/3},
\end{align}
which is shown by the green dashed curve in \Fig{rs.fig}. Even at full
gas conversion into stars, \hii{} regions are not well resolved for
$\nh \ga 10 \ \cci$, regardless of the resolution.

So far we assumed that the \stromgren{} sphere is powered by a single
stellar particle. The situation changes when it becomes likely to have
multiple young ($\la 5$ Myr) particles in a single cell, increasing
the source luminosity and the size of the \hii{} region. From
\Eeq{SFR.eq}, we can derive the star formation rate of a cell:
\begin{align}
  \SFRdx &= 1.75 \times 10^{-6} \ \msy \nonumber \\
  & \times \left( \frac{\dx}{18 \ \pc}  \right)^{3}
  \left( \frac{\sfeff}{0.02} \right)
  \left( \frac{\nh}{10 \ \cci}  \right)^{3/2},
\end{align}
from which we can then derive the hydrogen number density at which
more than one stellar particle, on average, is formed over $5$ Myr,
the time during which the stellar particles are luminous. This
density, at which we can start expecting multiple young stellar
particles per cell, is
\begin{align}
  \nh^{\rm mult} = 167 \ \cci  
  \left( \frac{\mstar}{600 \Msun}\right)^{2/3}
  \left( \frac{\sfeff}{0.02}  \right)^{-2/3}
  \left( \frac{\dx}{18 \ \pc}  \right)^{-2},
\end{align}
which coincides roughly with the density at which the track of
temperature versus density widens to the right, in the bottom phase
diagrams of \Fig{Ph_G9.fig}. However, the phase diagrams demonstrate
that, at the current resolution, the presence of multiple stellar
particles in a single cell is insufficient to resolve \stromgren{}
spheres at high densities, as the cells with multiple stellar
particles clearly do not reach $T\approx 10^4$ K.

We now see that \hii{} regions \emph{cannot} be fully resolved above
these moderately large gas densities, unless changes are made to the
star formation recipe, where e.g. more massive stellar particles are
formed from gas in a group of neighbouring cells or they are allowed
to accrete gas in their lifetime. Stellar particle accretion is
usually applied in numerical simulations of protostar formation and
the evolution of individual molecular clouds, i.e. simulations at
sub-galactic scales, and perhaps we are approaching a level of detail
which requires some merging of methods for these different scales of
galaxy evolution. Alternatively, one could apply stochastic radiation
feedback, by allowing on average one in X particles to emit radiation
at X times the default luminosity. Such an approach has been used by
\cite{DallaVecchia:2012br} and \cite{Roskar:2014be} for SN feedback to
overcome a related resolution problem of overcooling and an
over-smooth distribution of stars. Stochastic radiation and SN
feedback would thus appear to mesh quite naturally together to
overcome resolution problems.  This is beyond the scope of the present
paper though, and we can merely note the limitations in our feedback
at high densities, which are in any case close to the resolution
limit, where the pressure becomes dominated by the Jeans resolving
pressure floor. It is presently unclear what the exact effect of
under-resolved \hii{} regions is in our simulations, but likely it
leads to an underestimate of the effect of photoionisation heating,
since the gas in under-resolved \hii{} regions is heated to an
unrealistically low temperature (a fraction of the photo-ionised
temperature which corresponds roughly to the ratio of the size of the
real \hii{} region and the cell size).

\subsection{Analytic comparison of the RT feedback processes}
Among the main results of our simulations is that photoionisation
heating has a modest effect on regulating star formation, while
radiation pressure contributes negligibly. We now seek to understand
these results analytically, in order to see if they make physical
sense and to ensure that they are not the product of implementation
bugs.

We can compare, within our numerical framework, the efficiencies of
the different radiation feedback processes, i.e. photoionisation
heating, direct pressure from ionising photons, direct pressure from
optical photons, and multi-scattering pressure from IR photons. To
simplify and quantify this comparison, we consider feedback in a
single cell containing a radiation source and ignore radiation
entering the cell from the outside. While we will write the following
equations in terms of the simulation cell size, $\dx$, and the mass of
stellar particles, $\mstar$, most of the equations also hold
approximately for gas at a distance $\sim \dx$ from a star (cluster)
of mass $\mstar$ with the assumed (and theoretically motivated)
specific luminosity, provided the density, temperature, and
metallicity are nearly uniform within $\dx$.

We compare the radiation feedback efficiencies in terms of approximate
``effective'' temperatures. For photoionisation heating, this is equal
to the temperature of gas photoionised by stars, while for direct and
IR radiation pressure (with the motivation of comparing those
processes in a simple way), it is defined by equating the radiation
pressure and the thermal pressure.

\subsubsection{Photoionisation heating}
Photoionisation tends to heat the ionised gas to
$T_{\hii}\approx 2 \times 10^4$ K, as we have seen in
\Fig{Ph_G9.fig}. If the \stromgren{} radius extends outside the cell,
then the cell is simply heated to $T_{\hii}$, otherwise it is heated
to a fraction of that temperature which reflects the ratio of the
volume of the \stromgren{} sphere to that of the cell, i.e. the host
cell is heated to
\begin{align} \label{Teff_PH.eq}
  T^{\rm PH} \sim 2 \times 10^{4} \ {\rm{K}}
  \times {\rm min} \left( f_{\rm{vol}}, \ 1 \right), 
\end{align}
where 
\begin{align} \label{fvol.eq}
  f_{\rm{vol}} &= \frac{\frac{4}{3} \pi \rS^3}{\left( \dx \right)^3}
  &= \ 6.7 \
  \frac{\LumNumSpec_{\rm UV}}{5 \times 10^{46} \ \sm \Msun^{-1} } 
    \frac{\mstar}{600 \ \Msun} \\ 
    &&\times \left( \frac{\nh}{10 \ \cci} \right)^{-2}
  \left( \frac{\dx}{18 \ {\rm{pc}}} \right)^{-3},
  \nonumber
\end{align}
and we substituted our (\dw{} and \sbc{}) simulation parameters for
$\dx$, $\LumNumSpec_{\rm UV}$, and $\mstar$ (and the star formation
threshold for $\nh$). The specific stellar population luminosity, for
the UV and for the other photon groups, can be read (approximately)
from \Fig{groups.fig}.

\subsubsection{Direct pressure from photoionisation}
To quantify the effect of radiation pressure and compare it to
photoionisation heating, we measure it in terms of an effective
temperature, corresponding to the pressure applied via momentum
absorption from the radiation, and defined as
\begin{align}
  \Tnt \equiv \frac{P_{\rm rad} \mp}{\rho \kb}.
\end{align}
The radiation pressure is roughly the momentum absorption rate in the
cell, $\dot{p}$ (momentum per unit time), divided by the cell area,
\begin{align} \label{iso_P.eq}
  P_{\rm rad} = \frac{\dot{p}}{6 \left( \Delta x \right)^2}.
\end{align}
The momentum absorption rate can be estimated from the luminosity of
the stars contained in the cell and the opacity of the cell gas.  The
effective temperature is approximate, because we neglect the
dependence on the mean gas particle mass $\mu$, and we assume the
radiation pressure to be isotropic.

We assume for simplicity that the cell gas is in photoionisation
equilibrium with the emitted radiation\footnote{We ignore the
  instantaneous pressure from the radiation when the cell is in the
  process of being ionised, leading us to underestimate the direct
  pressure at low densities, where growing \hii{} regions are
  resolved.}, and we ignore the radial dependence of the neutral
fraction inside resolved \hii{} regions.  We can then use the size of
the predicted \hii{} region, given by \Eeq{rs.eq}, to estimate the
fraction of the ionising luminosity contributing to the direct
radiation pressure in the emitting cell, giving
\begin{align} \label{Teff_direct.eq}
  \Tnt^{\rm UV} &\sim \frac{L_{\rm UV}}{c}
                  \frac{1}{6 \left( \dx \right)^2}
  \frac{\mp}{\rho \kb} 
  \times {\rm min} \left( f^{-1}_{\rm{vol}}, \ 1 \right) \\ 
  &\sim 1.2 \times 10^3 \ {\rm K} \ \  
  \frac{\LumSpec_{\rm UV}}{2 \times 10^{36} \ \ergs \, \Msun^{-1} } 
    \frac{\mstar}{600 \ \Msun } \nonumber \\
    & \times \left( \frac{\nh}{10 \ \cci} \right)^{-1} 
  \left(\frac{\dx}{18 \ {\rm pc}} \right)^{-2} \times  {\rm min}
      \left( f^{-1}_{\rm{vol}}, \ 1 \right) \nonumber,
\end{align}
where we now measure the luminosity (and specific), $L_{\rm UV}$
($\LumSpec_{\rm UV}$), in terms of energy rather than photon count
(the value is again typical for the SED model utilised). The rightmost
term accounts for whether the \hii{} region is resolved or not: the
fraction of the ionising luminosity pressurising the emitting cell is
the volume of the \hii{} region over that of the cell, but this
fraction is roofed at unity, meaning all the emitted photons are
absorbed in the emitting cell as the \hii{} region becomes unresolved.

\subsubsection{Pressure on dust from optical photons}
The effective temperature corresponding to the pressure on gas via
dust from optical photons is
\begin{align}
  \Tnt^{\rm Opt} &\sim \frac{L_{\rm Opt}}{c}
      \frac{1}{6 \left( \dx \right)^2}
  \frac{\mp}{\rho \kb} \ \ 
  \left( 1-e^{-\tau_{\rm Opt}} \right) \\
  &\sim 1.8 \times 10^3 \ {\rm K} \ \  
  \frac{\LumSpec_{\rm Opt}}{3 \times 10^{36} \ \ergs \ \Msun^{-1} } 
    \frac{\mstar}{600 \ \Msun } \nonumber \\
  & \times  \left( \frac{\nh}{10 \ \cci} \right)^{-1} 
   \left(\frac{\dx}{18 \ {\rm pc}} \right)^{-2}
    \left( 1-e^{-\tau_{\rm Opt}} \right), \nonumber
\end{align}
where $\tau_{\rm Opt}$ is the optical depth of the host cell:
\begin{align} \label{tau_cell_UV.eq}
  \tau_{\rm Opt} &\sim \kOpt\rho\dx \\
  &\sim 1.2 \
  \frac{\kOOpt}{10^3 \ \ccg} \frac{Z}{\Zsun} \frac{\nh}{10 \ \cci}
  \frac{\dx}{18 \ {\rm pc}}. \nonumber
\end{align}

We ignore pressure on dust from UV photons, because the pressure from
the UV photons is already counted, in $\Tnt^{\rm UV}$, for
photoionisation, for which the opacity is orders of magnitude higher
than for dust absorption.

\subsubsection{Multi-scattering pressure on dust from IR photons}
The effective temperature for multi-scattering reprocessed IR photons
is
\begin{align} \label{Teff_multi.eq}
  \Tnt^{\rm IR} 
  &\sim \frac{L_{\rm IR}}{c} \frac{1}{6 \left( \dx \right)^2}
    \frac{\mp}{\rho \kb} \tauIR
    = \frac{L_{\rm IR}}{c}\frac{\kappa_{\rm IR}
    \mp}{6 \dx \kb} \\
  &\sim 22 \ {\rm K} \  
  \frac{\LumSpec_{\rm Opt}}{3 \times 10^{36} \ \ergs \ \Msun^{-1} }
    \frac{\mstar}{600 \Msun } \nonumber \\
  & \times  \frac{\kOIR}{10 \ \ccg}
    \frac{Z}{\Zsun}
  \left( \frac{\dx}{18 \ {\rm pc}} \right)^{-1}. \nonumber
\end{align}
We only consider the optical stellar luminosity, since the IR
luminosity is negligible in comparison (see \Fig{groups.fig}). The IR
multi-scattering feedback depends on the optical photons being
absorbed and re-emitted into the IR. It is a safe assumption though,
that this is true under any circumstances where multi-scattering is
important, since $\kappa_{\rm IR} \ll \kappa_{\rm Opt}$.

The expression for the IR effective temperature assumes trapping of
photons originating \emph{within} the cell and ignores additional
trapping of photons originating from the neighbouring environment. The
previous expression should therefore be taken as a lower limit. This
is less of a concern for the other radiation feedback processes, since
they are unresolved at high (star-forming) densities, and we thus
expect much less inter-cell flux of photons.

\subsubsection{Relative impact of the radiation feedback processes}
To compare the radiation feedback processes
(Eqs. \ref{Teff_PH.eq}-\ref{Teff_multi.eq}), we replace the stellar
mass, $\mstar$, by the fraction $f_* \le 1$ of the gas mass in a cell
at a given density and volume. The value $f_*=1$ gives an approximate
upper limit on the effective temperature estimates for each of the
processes, with the possible exception of the IR radiation, where
things are more uncertain, as argued in the previous sub-section. We
thus assume
\begin{align} \label{mstar.eq}
  \mstar =f_* \frac{\nh \mp}{X} (\dx)^3 =
  1875 \ \Msun  \ f_* \frac{\nh}{10 \, \cci} 
  \left( \frac{\dx}{18 {\rm pc}} \right)^3.
\end{align}
In practice, the mass of each stellar particle is some fraction ($1/3$
in the \dw{} and \sbc{} simulations, $2/3$ in \mw{}) of the cell gas
mass at the density threshold for star formation. The upper limit we
use in \Eeq{mstar.eq}, reflects the fact that at large densities,
multiple stellar particles can be expected to form in the same cell
over a short timescale, but no more than the total gas mass in a cell
can be converted into stars. 

Substituting \Eeq{mstar.eq} into
(\ref{Teff_PH.eq})-(\ref{Teff_multi.eq}) gives:
\begin{align}
  T^{\rm PH} &= 2 \times 10^{4} \ {\rm{K}}
  \times {\rm min} \left( f_{\rm{vol}}, \ 1 \right), 
  \label{Teff_PH_gen.eq} \\
  \Tnt^{\rm UV} &= 3.7 \times 10^3 \ {\rm K} \  
  \frac{\LumSpec_{\rm UV}}{2 \times 10^{36} \ \ergs \ \Msun^{-1}} 
  \frac{\dx}{18 \ {\rm pc}} \nonumber \label{Teff_UV_gen.eq}  \\
  &\times  {\rm min} \left( f^{-1}_{\rm{vol}}, \ 1 \right) 
    \ f_*,
  \\
  \Tnt^{\rm Opt} &= 5.6 \times 10^3 \ {\rm K} \  
  \frac{\LumSpec_{\rm Opt}}{3 \times 10^{36} \ \ergs \ \Msun^{-1} } 
  \frac{\dx}{18 \ {\rm pc}}  \label{Teff_Opt_gen.eq} \\
  &\times   
  \left( 1-e^{-\tau_{\rm Opt}} \right) \ f_*, \nonumber
  \\
  \Tnt^{\rm IR} &= 69 \ {\rm K} \  
  \frac{\LumSpec_{\rm Opt}}{3 \times 10^{36} \, \ergs \, \Msun^{-1}} 
  \frac{\kOIR}{10 \ \ccg}
  \frac{Z}{\Zsun } \label{Teff_IR_gen.eq} \\
  & \times
  \frac{\nh}{10 \, \cci }
  \left( \frac{\dx}{18 \ {\rm pc}} \right)^2 \ f_*, 
  \nonumber
\end{align}
where the volume fraction,
\begin{align} \label{fvol_general.eq} 
  f_{\rm{vol}} = \left( \frac{\nh}{2.1 \times 10^2 \, \cci} \right)^{-1}
  \frac{\LumNumSpec_{\rm UV}}{5 \times 10^{46} \, \sm \, \Msun^{-1} } 
  \ f_*,
\end{align}
is independent of resolution.

\begin{figure}
  \centering
  \includegraphics
    {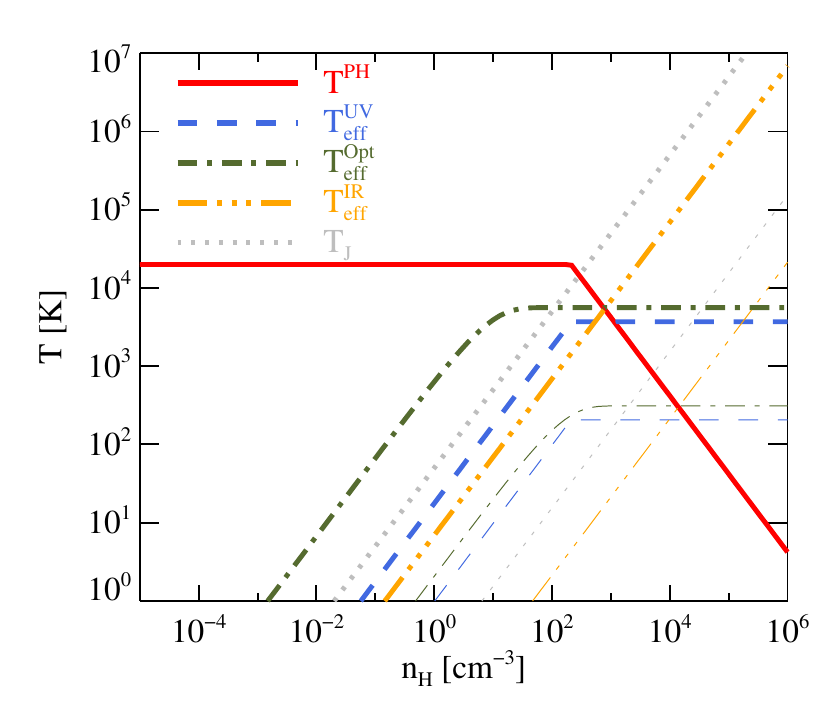}
  \caption
  {\label{rtfb_channels.fig}Effective temperatures acquired in a
    single emitting cell via the different radiation feedback
    processes (Eqs. \ref{Teff_PH_gen.eq}-\ref{Teff_IR_gen.eq}),
    assuming $Z=\Zsun$ and that a fraction $f_*=1$ of the cell gas is
    converted instantaneously into stars. For the thick lines, we
    assume $\dx=18$ pc, while for the thin lines we assume $1$ pc
    resolution. The processes considered are: photoionisation heating
    (solid red), direct pressure from photoionisation (dashed blue),
    direct pressure via dust from optical photons (dot-dashed green),
    and multi-scattering pressure from IR photons (dot-dot-dot-dashed
    yellow). Also plotted, in dotted grey, is the effective
    temperature of the density-dependent pressure floor
    (\Eq{polytrope.eq}). Photoionisation heating dominates the
    radiation feedback at low densities in this single cell limit,
    while the pressure floor takes over at high densities.  The
    first-order effect of increasing the resolution is to decrease the
    effect of radiation pressure. However, at extreme densities, IR
    trapping by multiple stellar particles on scales larger than the
    cell width is likely to give a boost over that indicated in the
    plot.}
\end{figure}

These effective temperatures are plotted in \Fig{rtfb_channels.fig},
for $f_*=1$ and $Z=\Zsun$, with the thick curves representing the $18$
pc resolution used for our lower-mass galaxies, and the thin curves
corresponding to $\dx=1$ pc. We also plot (in dotted grey) the
artificial non-thermal pressure, \Eeq{polytrope.eq}, that is imposed
to resolve the Jeans scales, and we shift it with resolution according
to \Eeq{Tfloor.eq}.

Photoionisation heating (solid red) dominates at low densities,
$\nh \la 10^2 \, \cci$, heating the gas to $\approx 2 \times 10^4$ K,
while for higher densities, radiation feedback is surpassed by the
Jeans pressure. In the absence of the Jeans pressure, IR
multi-scattering would dominate at high densities. The direct UV and
Optical radiation effective temperatures plateau at high densities, as
the total particle luminosity becomes absorbed in the local cell, and
increasing the resolution only makes radiation pressure weaker, since
the lower stellar mass (and hence luminosity) has a stronger negative
effect ($\propto (\dx)^3$) compared to the positive effect of the
decreased cell area ($\propto (\dx)^{-2}$). The IR radiation pressure
is the only term which keeps rising for increasing densities, which is
due to the multi-scattering, and it dominates over other radiation
feedback processes at extreme densities. The IR effective temperature
is, however, lower than the artificial pressure floor, $\TJeans$, by
about an order of magnitude (at $Z=\Zsun$), at any resolution.

\Fig{rtfb_channels.fig} qualitatively justifies the results of our
simulations. Comparing the plot to the bottom right phase diagram of
\Fig{Ph_G9.fig}, the plateau of $\approx 2 \times 10^4$ K gas is clear
in both figures, and the drop-off in temperature, which is a
manifestation of unresolved \hii{} regions, occurs at a similar
density, though slightly lower in the phase diagram, which is because
less than the full cell mass is converted into stars in the
simulations ($f_*=1/3$).  Radiation (heating) feedback is effective in
preventing gas at low densities from clumping, but futile in
dispersing clouds once the densities become high. Radiation pressure
vanishes at low densities, and is negligible compared to the
artificial Jeans pressure at high densities.

Judging from \Fig{rtfb_channels.fig}, it appears that the modelled
radiation pressure is doomed to always remain weaker than the
artificial pressure floor we are forced to apply, especially
considering that we are assuming an extreme upper limit where the full
gas mass in a cell is converted instantaneously into stars.  However,
we stress again that we only consider in this analysis the effect in a
single cell, and ignore the effect appearing from external stellar
populations in neighbouring cells (and we also ignore the fact that
particles can move to higher or lower densities during their
lifetime). We can conclude that direct radiation pressure is weak at
any resolution, but with many stellar particles forming in highly
resolved optically thick regions, we may see a considerable boost in
the pressure from trapped multi-scattering IR radiation with higher
resolution (and only at high densities). It remains a task for future
work to establish what kind of resolution is required to see such a
boost, but in the next subsection we will investigate what effects we
can expect from it on large scales.

The above analysis does not apply to the time-integrated effect of
collective, direct long-range radiation pressure from many stellar
populations on galactic scales, e.g. in stirring diffuse gas or
pushing cold clouds out of the galaxy \citep{Murray:2011en}. However,
this effect is unimportant in our simulated galaxies, since it exists
(and is in fact exaggerated due to our full reduced flux
approximation), yet radiation feedback is dominated by heating, with
radiation pressure at best having a marginal impact.

\subsection{What does it take for IR radiation pressure to dominate,
  and what happens then?}\label{whatif.sec}
Up to this point, we have found that IR radiation has only a marginal
effect on our galaxies, and we have shown analytically that these
results are to be expected with our current model and resolution. The
considerable increase in resolution, that appears to be required to
investigate IR pressure feedback on small scales, and possible
cascading effects on larger scales, is beyond our reach in the current
paper. We can, however, instead artificially allow the IR radiation to
dominate the galactic feedback by simply increasing the IR
opacity. This can give us an estimate of how far we are from efficient
IR feedback, and, more importantly, how the galaxy reacts when IR
feedback does become efficient on small scales. Thus we get an idea
about what to expect \emph{if} we resolve the very large optical
depths that are required for multi-scattering radiation pressure to
play a role. For example, does the radiation generate large-scale
winds, and does it create a much thicker gas or stellar disk? We thus
ran variants of the \mw\_\simf{} simulation (where the IR optical
depths are largest) with increased IR opacities. We compare here
results for $\kOIR=(10^2, 10^3, 10^4) \, \ccg$, i.e. ten, a hundred
and a thousand times the default, physically motivated, opacity that
we have used so far. We also set $\kOUV=\kOOpt=\kOIR$ in the highest
opacity run in order to increase the IR reproduction from
higher-energy photons in line with the opacity increase.

\begin{figure}
  \centering
  \includegraphics[width=0.47\textwidth]
    {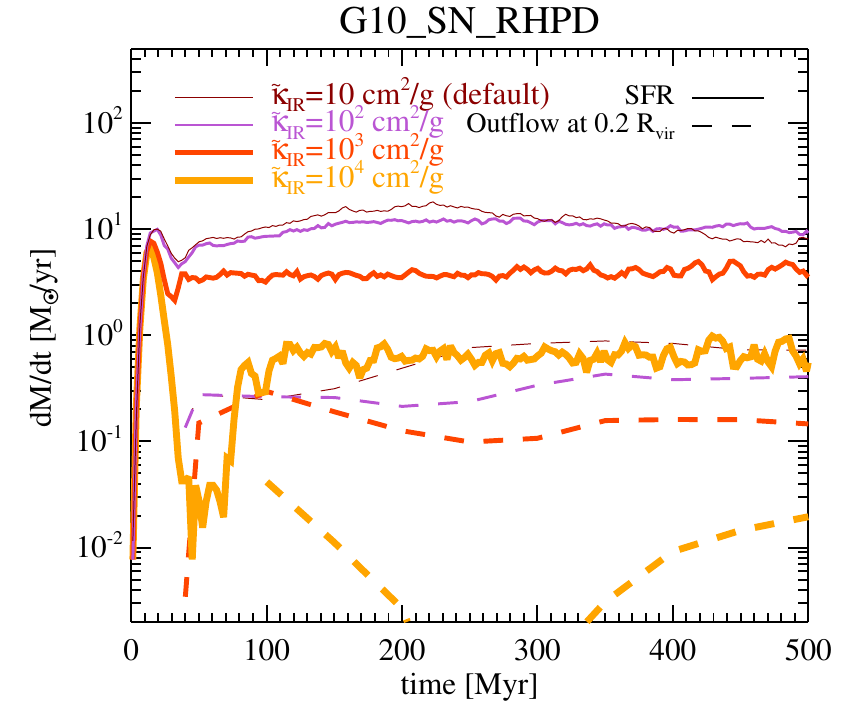}
  \caption
  {\label{sfr_G10_kvar.fig}Star formation rates (solid lines) and
    outflow rates across planes at distances of $0.2 \ \Rvir$ from the
    disk plane (dashed lines) in the \mw\_\sime{} galaxy with
    increased IR opacity. The thinnest (dark red) curve shows the
    results for the default opacity that we have used so far and the
    successively thicker curves show results where the IR opacity is
    increased, each time by a factor of ten. The star formation
    becomes bursty in the case with the highest opacity. Outflow rates
    decrease with increasing opacity, more or less in line with the
    reduced star formation, indicating that the radiation disrupts
    star-forming clouds gently, rather than violently.}
\end{figure}

\Fig{sfr_G10_kvar.fig} shows star formation rates and outflow rates
across planes $0.2 \ \Rvir$ from the disk plane. We hardly see any
effect on the star formation rate, though the outflow rate is slightly
reduced. Further increased opacity increasingly suppresses both star
formation and outflow rates, with a very bursty star formation and
almost totally quenched outflows in the most extreme case. The
reduction in the outflow rate, which is at least as large as for the
SFR, even with the appearance of bursty star formation, strengthens
the impression of a non-violent radiation feedback, which stirs up the
gas but does not systematically eject it.

\begin{figure}
  \centering
  \includegraphics[width=0.45\textwidth]
    {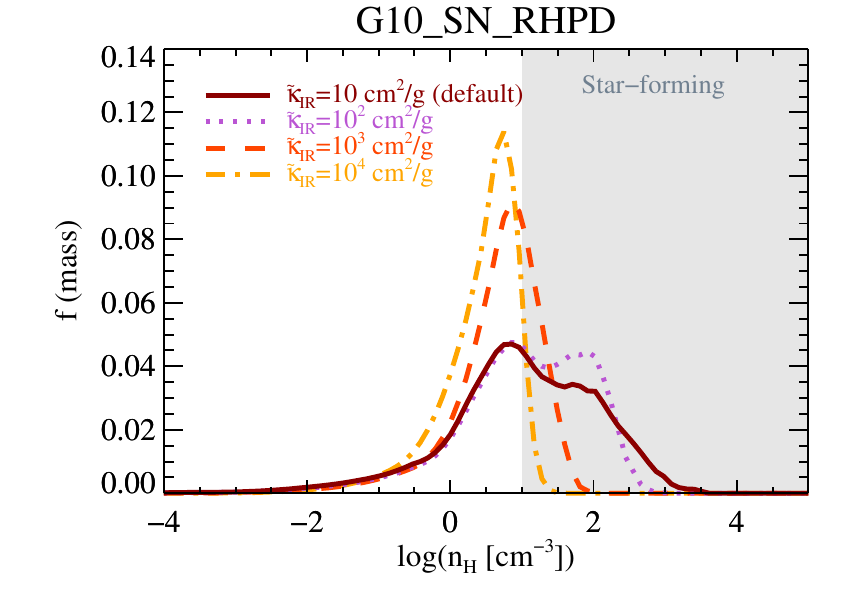}
  \caption
  {\label{hist_nH_G10_kvar.fig}Time-stacked mass-weighted gas density
    distribution in the \mw\_\sime{} galaxy, with increasing IR
    opacity. For the highest opacity used, when the star formation is
    reduced by more than an order of magnitude, the density
    distribution becomes cuts off at the star formation threshold,
    $\nh = 10 \, \cci$.}
\end{figure}

\Fig{hist_nH_G10_kvar.fig} shows the effect of the increased opacities
on the gas density distribution. As expected, the IR radiation
suppresses high densities more efficiently with increased opacity. For
the highest opacity, the density distribution cuts off at the
star-formation density threshold ($n_{*}=10 \ \cci$). This indicates
that the IR pressure does indeed become very efficient at preventing
gas to form stars, but that is more or less the whole effect, i.e. the
gas is kept diffuse, but within the galaxy. The IR pressure shifts SN
explosions to lower densities, but this does not lead to increased
outflow rates, as the reduced star formation more than compensates to
reduce the outflows, leading to a decrease in mass loading with
increasing opacity.

\begin{figure}
  \centering
  \subfloat
  {\includegraphics[width=0.47\textwidth]
    {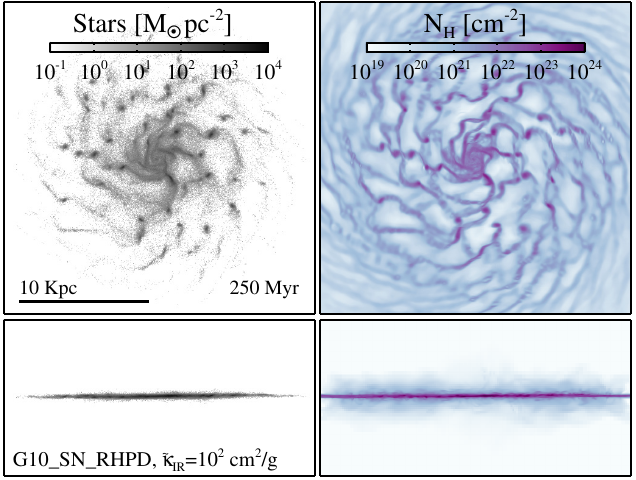}}
  \vspace{1.mm}
  \subfloat
  {\includegraphics[width=0.47\textwidth]
    {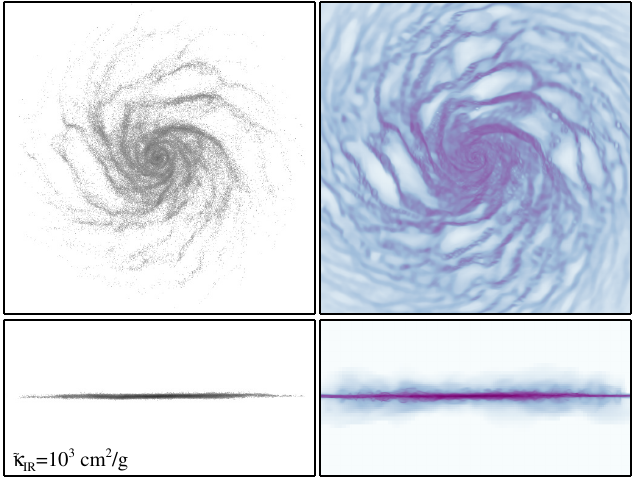}}
  \vspace{1.mm}
  \subfloat
  {\includegraphics[width=0.47\textwidth]
    {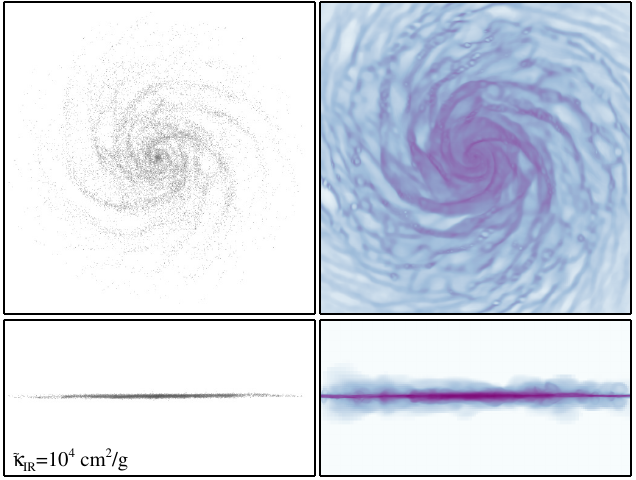}}
  \caption
  {\label{maps_G10_kvar.fig} Maps of the \mw{} galaxy
    ($3.5 \times 10^{10} \ \Msun$ in baryons) at $250$ Myr, for SN and
    full RT feedback (same as bottom left panel in
    \Fig{maps_G10.fig}), but with increased IR opacity, as indicated
    in each panel: the opacities are increased from the default by a
    factor $10$ (top), $10^2$ (middle), and $10^3$ (bottom). The
    increased dominance of IR radiation pressure simply has the effect
    of smoothing out the peaks in gas density, and suppressing star
    formation.}
\end{figure}

Finally, we compare the galaxy morphologies for the different opacity
values in \Fig{maps_G10_kvar.fig}. The galaxy simply becomes smoother
with increasing opacity, both in the gas and stars.

\subsection{Comparison with other work} \label{comparison.sec}

While many studies exist in the literature where radiation feedback
(on galactic scales) is modelled with subgrid recipes in pure HD
simulations, there are only a few in which a subset of the radiation
feedback mechanisms that are modelled here are studied with RHD
\citep{Petkova:2011jh,Wise:2012dh,Kim:2013ie,Hasegawa:2013fv,Pawlik:2013jz,
  Pawlik:2015vx}.

In cosmological simulations of reionisation-era galaxy formation,
\cite{Wise:2012dh} found direct radiation pressure to play a major
role, suppressing star formation strongly and boosting outflow rates,
but did not report on the isolated effect of photo-ionisation heating,
which was always included, and they did not include IR radiation
effects (which are likely weak in such metal-poor
galaxies). \cite{Petkova:2011jh, Hasegawa:2013fv, Pawlik:2013jz} also
modelled reionisation-era galaxies, but only included radiation
heating. They all found that radiation heating gently suppressed star
formation, while they did not report on boosted outflows.
\cite{Pawlik:2015vx} considered the addition of SN feedback and found
that it dominated over the effect of radiation heating on star
formation histories. \cite{Kim:2013ie} used the RHD implementation
from \cite{Wise:2012dh} on an isolated $\sim$ MW mass galaxy, and
found slight ($\sim 20\%$) suppression of star formation, which they
attributed to radiation heating, rather than radiation pressure.

A large amount of work exists where radiation feedback has been
included in pure HD simulations in the form of subgrid recipes, which
are often empirically motivated. Although there are quantitative, and
sometimes qualitative, differences, these studies broadly agree that
IR radiation pressure strongly suppresses star formation and generates
(sometimes extremely massive) outflows \citep[e.g.][]{Stinson:2013ex,
  Hopkins:2011fk, Hopkins:2012hr, Hopkins:2012ez, Hopkins:2012bm,
  Aumer:2013fm, Agertz:2013il, Agertz:2015gb, Roskar:2014be}. There
are exceptions though: \cite{Ceverino:2014dw} and \cite{Moody:2014gk}
found direct radiation pressure to mildly suppress star formation,
while radiation heating and IR radiation pressure had a negligible
effect.  \cite{TrujilloGomez:2015dp}, on the other hand, found
radiation heating dominated over direct radiation pressure in
suppressing star formation, while outflows were not affected by the
radiation (and IR effects were not considered).

Our results do not show a wide and general agreement with previous
studies of the effects of radiation feedback on galactic scales, which
is not surprising, since there is no general agreement in the
literature. The discrepancies probably largely come down to
resolution. It appears that both RHD and HD simulations that show a
substantial effect from IR radiation pressure have either sub-pc
resolution or a subgrid model that boosts the optical depths. We lack
sub-pc resolution in the current paper, and we have so far made no
attempt to compensate this with a subgrid model. Of these two options,
we prefer in future work to increase the resolution, to probe from
first principles how radiation feedback affects small scales, and how
this effect may (or may not) cascade to larger scales. The strongest
general disagreement we can find with other work concerns
outflows. Where they are studied in the literature, radiation feedback
appears to boost outflows most of the time, which is in contrast with
our simulations. Our experiments with boosted IR radiation opacities
hint that increased resolution will still leave us with a lack of
radiation-generated outflows, but in the end, the best way to find out
is to actually increase the resolution.

\section{Conclusions and future work} \label{Conclusions.sec} 

We ran and analysed adaptive mesh refinement simulations of isolated
disk galaxies of baryonic masses
$3.5 \times (10^{8}, 10^{9}, 10^{10}) \, \Msun$ (the largest mass
being comparable to that of the Milky Way), using a maximum resolution
of $18$ pc. We studied the effects or stellar radiation feedback,
which was modelled with radiation-hydrodynamics, acting on its own and
also combined with (``thermal dump'') supernova feedback. We compared
the effects of three separate radiation feedback processes:
photoionisation heating, direct radiation pressure from UV and optical
photons, and pressure from multi-scattered, reprocessed IR
radiation. These are the first galaxy-scale simulations which model
all these processes concurrently and with RHD. Our main findings are
the following:

\begin{itemize}
\item Stellar radiation feedback suppresses star formation in the
  simulated galaxies. It does so predominantly by preventing the
  formation of star-forming clumps, rather than by destroying those
  that form. The suppression of star formation with radiation feedback
  (ranging from a factor of 4 for the low mass galaxy to only
  $\sim 0.1$ for the most massive one) is similar to that of ``thermal
  dump'' SN feedback.
\item Radiation feedback does not significantly amplify the efficiency
  of SN feedback, and in fact there is a hint of the opposite effect
  in the lowest-mass galaxy we consider, where the combination of
  radiation and SN feedback results in a weaker star formation
  suppression than one would naively expect from multiplying the
  individual suppression factors, although the combined effect does
  exceed that of the individual feedback processes.
\item Radiation feedback has a negligible effect on galaxy outflows.
  If anything, the outflow rates are slightly suppressed, owing to the
  reduced star formation and subsequent decrease in SN activity. 
  The outflow mass loading factor, i.e. the ratio between the outflow
  rate and the star formation rate, is typically of the order of
  $10^{-1}$, which is very low compared to non-RHD simulations that
  use subgrid recipes for radiation feedback
  \cite[e.g.][]{Hopkins:2012ez}.
\item As with (``thermal dump'') SN feedback, the effect of radiation
  feedback on star formation weakens with galaxy mass and
  metallicity. The combined effect of SN and radiation feedback is
  strongest in our intermediate-mass galaxy, which has a baryonic mass
  of $3.5 \times 10^{9} \Msun$, i.e. about one-tenth of the mass of
  our Milky-Way.
\item The dominant form of radiation feedback is photoionisation
  heating, while the effect of radiation pressure, both direct and on
  dust, is borderline negligible. We are able to explain the relative
  efficiencies of the different radiation feedback processes using
  simple analytic estimates within the context of our numerical
  models.
\item The analytic estimates suggest that the effect of direct
  radiation pressure from ionising radiation on galaxies is likely
  negligible in reality. However, multi-scattering radiation pressure
  from IR radiation is not properly captured in our simulations. This
  is because our resolution ($\sim 10$ pc) does not allow the collapse
  to sufficiently large local densities for the gas to become
  significantly optically thick to the IR radiation.
\item To estimate the qualitative effects of multi-scattering IR
  radiation that may be revealed by future, higher-resolution,
  simulations, we ran simulations in which the IR opacities were
  boosted by orders of magnitude over realistic, physically motivated,
  values. This boost makes multi-scattering radiation effective at
  regulating star formation, but in a gentle way that merely smooths
  out the galaxy disk, without generating outflows.
\item Resolution is also an issue for the ionising radiation. With the
  current method for forming stars, where stellar population particles
  are instantaneously formed out of the gas in a single AMR cell,
  \hii{} regios are unresolved at densities $\nh \ga 10 \, \cci$,
  \emph{regardless of the numerical resolution}. The consequence of
  the unresolved \hii{} regions is likely an underestimate of the
  regulating effect of radiation heating on star formation in dense
  gas, since the gas cells hosting young stars are heated to
  temperatures lower than the ionisation temperature. Possible ways to
  deal with this problem in the future include stochastic radiation
  feedback or a modified method for star formation.
\item Although we have not considered this in detail, we find in the
  resolution tests described in \App{restests.sec} that radiation
  feedback is much less sensitive to the stellar particle mass than is
  (``thermal dump'') SN feedback. This makes sense, since the radiation
  is continuous, while the SNe are instantaneous, and for explosive
  feedback the radiative losses decrease for higher maximum
  temperatures.
\end{itemize}

An important caveat for our study is that our simulated galaxies do
not have the high surface densities that occur in the massive,
starbursting galaxies that have been the focus of theoretical work
which predicts efficient regulation of star formation and outflows by
radiation pressure \citep[e.g.][]{Murray:2011en}. At high-redshift
($z \sim 3$), where gas accretion and star formation peak, radiation
pressure may even play a role in `normal' low-mass galaxies.  In the
future we will expand our simulations to include more massive
galaxies, and gas-rich galaxies representative of high redshift, which
may exhibit greater sensitivity to radiation pressure.

We also note that the choice of SN feedback recipe likely affects the
interplay of feedback processes and the net effect of radiation
feedback. For simplicity, and in order to make sure we did not
over-inject feedback energy in this first round of simulations, we
used ``thermal dump'' SN feedback, which is known to be inefficient
and suffer from resolution-induced overcooling. In future studies it
will be interesting to see how the interplay of feedback processes is
affected by the use of more efficient (and more realistic) SN feedback
recipes.

There are many additional interesting paths to follow, such as
improvements of our radiation feedback model, the inclusion of other
sources of radiation than stars, and an expansion to both larger and
smaller physical scales.

An interesting model improvement is to consider the effect of the
local radiation field on metal cooling, which has been suggested by
\cite{Cantalupo:2010jw} to effectively quench cooling and subsequently
star formation in galaxies. Another important model improvement is the
inclusion of the formation and radiative dissociation of $H_2$, which
is highly relevant for studying star formation in detail. AGN feedback
may be fundamentally radiative in origin, and it is quite interesting
to see in what ways, if any, RHD experiments would differ from subgrid
recipes. It is relatively straightforward to add AGN radiation to our
simulations, as long as a recipe for black hole accretion is in
place. Some additional radiation processes, such as Compton
scattering, are likely important, and it is quite likely that it will
remain difficult to resolve optically thick regions properly.

We intend to study radiation feedback on scales both larger and
smaller than the current study. The larger scales involve cosmological
zoom RHD simulations, where the effect of radiation feedback can be
studied in galaxies that evolve in their natural environment, and we
can study the effect on galaxy evolution, inflows, outflows, and the
observable properties of the ISM and CGM. Going to smaller scales will
allow us to properly resolve optically thick star-forming clouds and
how they are affected by stellar.

\section*{Acknowledgements}
For discussions and suggestions contributing to this paper, we thank
Jeremy Blaizot, Benoit Commercon, Yohan Dubois, and Alireza Rahmati,
and we are grateful to the anonymous referee for constructive
comments. This work was funded by the European Research Council under
the European Union’s Seventh Framework Programme (FP7/2007-2013) / ERC
Grant agreement 278594-GasAroundGalaxies, and the Marie Curie Training
Network CosmoComp (PITN-GA-2009-238356). The simulations were in part
performed using the DiRAC Data Centric system at Durham University,
operated by the Institute for Computational Cosmology on behalf of the
STFC DiRAC HPC Facility (www.dirac.ac.uk). This equipment was funded
by BIS National E-infrastructure capital grant ST/K00042X/1, STFC
capital grant ST/H008519/1, and STFC DiRAC Operations grant
ST/K003267/1 and Durham University. DiRAC is part of the National
E-Infrastructure. Computer resources for this project have also been
provided by the Gauss Centre for Supercomputing/Leibniz Supercomputing
Centre under grant:pr83le. We further acknowledge PRACE for awarding
us access to resource Supermuc based in Germany at LRZ Garching
(proposal number 2013091919), and to the ARCHER resource
(http://www.archer.ac.uk) based in the UK at the University of
Edinburgh (PRACE-3IP project FP7 RI-312763).

\bibliography{ref}

\appendix

\section{Stellar luminosities and photon properties}\label{groups.app}

\begin{figure*}
  \centering
  \includegraphics[width=0.85\textwidth]
    {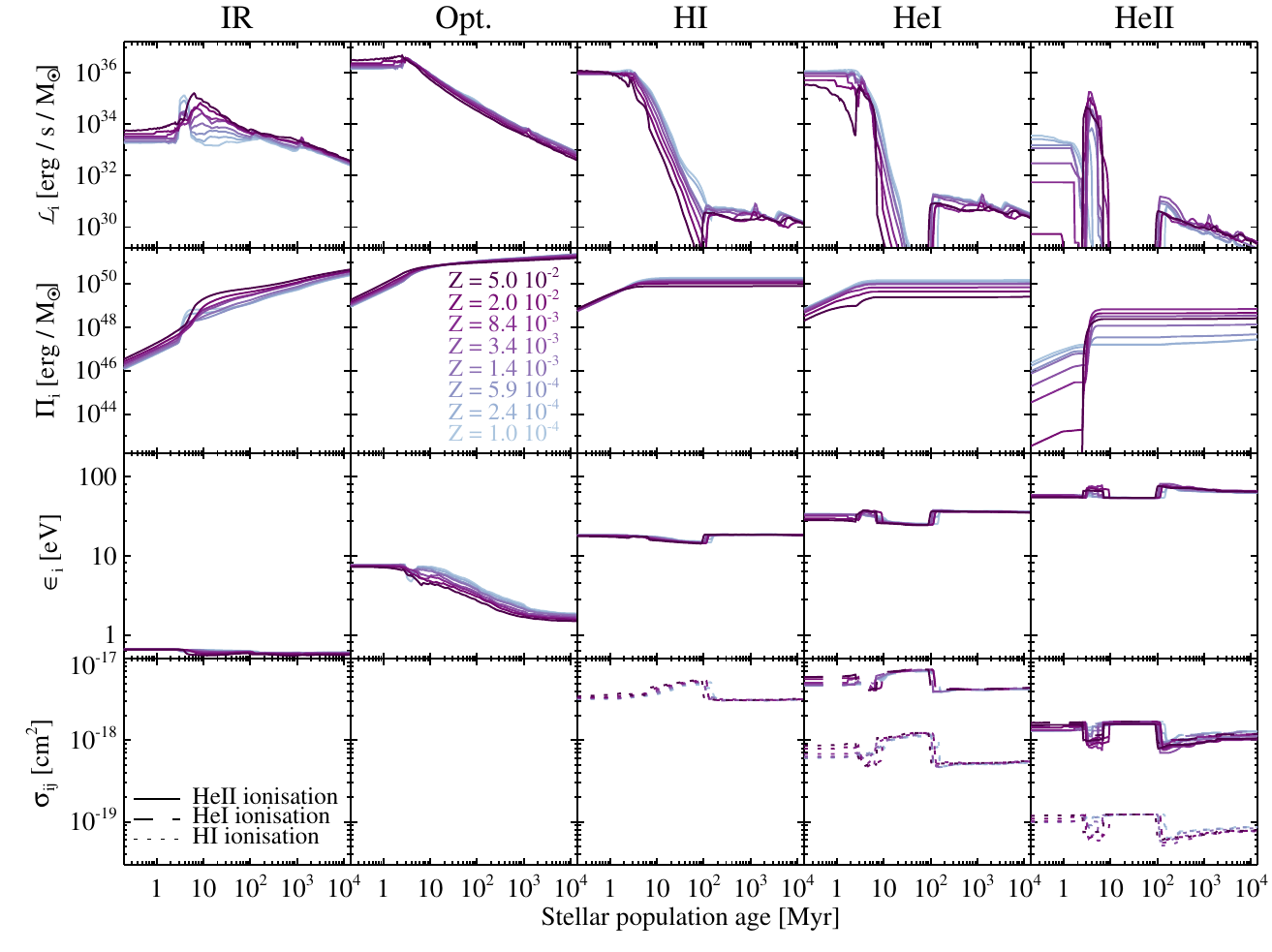}
  \caption
  {\label{groups.fig}Age and metallicity dependence of specific
    stellar luminosities and photon group attributes derived from the
    \protect\cite{Bruzual:2003ck} SED model, assuming a
    \protect\cite{Chabrier:2003kia} IMF. The columns represent the
    five photon groups with increasing photon energy from left to
    right. \textbf{Top row:} specific luminosity (i.e. luminosity per
    unit stellar mass), emitted into each photon group from the
    stellar particles.  \textbf{Second row:} cumulative specific
    luminosity per photon group. \textbf{Third row:} average photon
    energies. \textbf{Bottom row:} average cross sections per
    photoionisation interaction. The emission from each stellar
    particle is calculated on-the-fly in each timestep by integration
    of the data shown in the second row, given the mass, age and
    metallicity of the particle. The properties of the five photon
    groups are updated every five coarse timesteps by a
    luminosity-weighted average of all existing stellar particles
    (excluding the stellar particles present in the initial
    conditions).}
\end{figure*}

The emission from each stellar particle is calculated on the fly for
every fine timestep and injected into the host cell, adding to the
radiation energy density of all photon groups. For the specific
stellar luminosity (i.e. luminosity per unit mass) and photon group
properties, we use the spectral emission distribution (SED) models of
\cite{Bruzual:2003ck}, where we assume a \cite{Chabrier:2003kia}
IMF. The dependence of the specific luminosities and radiation group
properties on the stellar population's age and metallicity are shown
in \Fig{groups.fig}. Each photon group's properties (i.e. average
energy and cross section) are updated every 5 coarse timesteps, using
luminosity-weighted averages of the existing stellar particles'
emission in the corresponding bands using the frequency dependent
ionisation cross sections from \cite{Verner:1996dm} (see also the
appendix of \RRT{}). This update of the photon groups and the stellar
emission is done as detailed in the appendix of \RRT, except that the
specific luminosity is now in terms of emitted energy whereas it was
done in terms of photon number count in \RRT{} (which makes more sense
in pure ionisation calculations). The stellar emission is thus
energy-conserving, whereas it was photon number conserving in
\RRT. This difference arises because the spectral shape of an
individual stellar particle is not identical to the ``average'' shape
which is assumed for our photon groups. Due to this difference, one
must choose whether the emission is accurate in terms of photon count
or energy, and we have chosen energy.

\section{The reduced flux approximation}\label{redflux.app}
In \Sec{methods.sec}, we describe the reduced flux approximation,
whereby we assume a full reduced flux of photons, $|{\bf F}|=\cred E$,
when calculating the direct radiation pressure on gas from non-IR
photons. The reason for making this approximation is as
follows. Radiation is emitted from a stellar particle directly into
the cell which hosts the particle, by incrementing the radiation
energy density, $E$, while the photon flux is left unchanged, in
accordance with locally isotropic radiation from the stellar
population. We use the so-called Global Lax Friedrich Riemann (GLF)
solver for the advection of photons between cells (see \RRT{}, \RM{}),
which has the advantage that radiation is advected isotropically from
such sources, i.e. the radiation field retains an isotropic shape, in
the limit of a single source and free-streaming radiation. The
disadvantage of the GLF solver is that the radiation stays somewhat
isotropic inside a buffer of a few cell widths around such a source,
i.e. in this region $|{\bf F}| \ll \cred E$.

\begin{figure}
  \centering
  \includegraphics[width=0.47\textwidth]
    {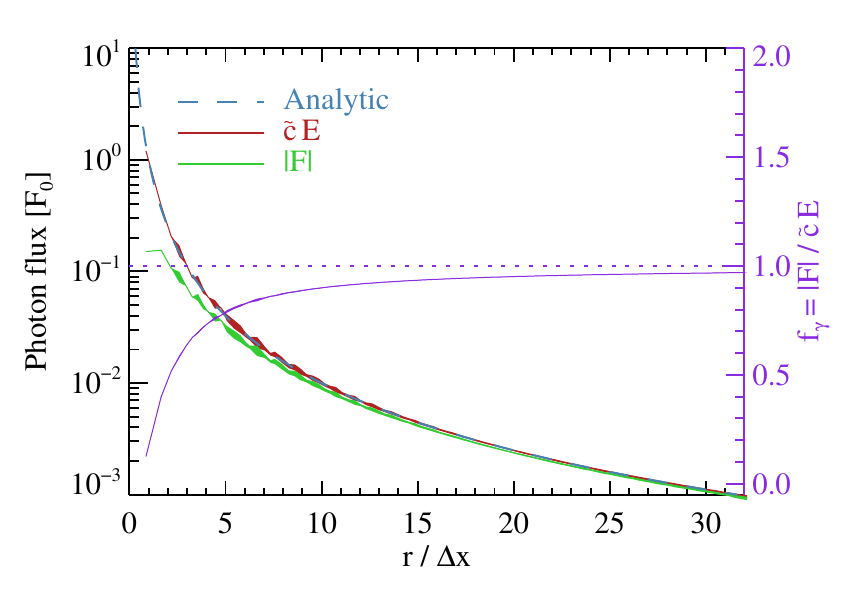}
  \caption
  {\label{redflux.fig}Converged results from a simple 3-d experiment
    of a single isotropic source of free-streaming radiation in the
    center of a box resolved by $64^3$ cells. We plot the analytic
    expectation for the $r^{-2}$ radiation flux (dashed blue), the
    ``angle-integrated'' radiation flux, $\cred E$ (solid red), and
    the magnitude of the radiation flux away from the source, $|F|$
    (solid green), all in units of $F_0$, the expected analytic flux
    at a distance of one cell width from the source. Against the right
    axis, in purple, we plot the reduced flux, $\fred$, which is the
    ratio of $|F|$ and $\cred E$. The reduced flux curve demonstrates
    a ``smoothing length'' for the reduced flux of a few cell widths
    around the isotropic radiation source, within which the radiation
    pressure, $\dot {\bf p}_{\gamma} \propto {\bf F}$, is considerably
    underestimated. The curves for $\cred E$, $|F|$, and $\fred$ have
    been binned by radius, and the thickness of the curves reflects
    the flux range within each radial bin.}
\end{figure}

We demonstrate this in \Fig{redflux.fig}, where we show the converged
results of a 3-d experiment of a single isotropic source of radiation
in the middle of a box resolved by $64^3$ cells, and assuming
free-streaming radiation, i.e. no interaction between the radiation
and the medium. The photons are injected by incrementing $E$ during
each timestep, uniformly in the eight cells adjacent to the box
center, according to the luminosity of the source. We plot, as a
function of distance $r$ from the source, the analytic expression for
the radiation flux, i.e. the ``angle-integrated'' flux $\cred E$, and
the magnitude of the photon bulk flux, $|F|$ (pointing away from the
source), all in units of $F_0=\frac{L}{4 \pi \Delta x^2}$, where $L$
is the source luminosity and $\Delta x$ the cell width. It can be seen
from the plot that $\cred E$ follows the analytic profile accurately,
but within a few $\Delta x$ from the source, $|F| \ll \cred E$. In
solid purple, against the right axis, we plot also the reduced flux of
the radiation, $\fred = \frac{|F|}{\cred E}$.  While in reality, one
would have $\fred=1$ at \emph{any} radius for this simple experiment,
this is clearly quite far from the truth close to the source, with
e.g. $\fred(r<5 \Delta x) \la 0.8$. For the advection of photons,
photoionisation and the associated heating, this is of no consequence,
but the \emph{radiation pressure} is correspondingly underestimated in
such a buffer of $\approx$ 5 cell widths around the source, which can
be considered a ``smoothing length'' for the radiation pressure
$\dot {\bf p}_{\gamma} \propto {\bf F}$ (see \RM{}, Sec. 2.3.3, and
Eq. 27). Typical \hii{} regions in our simulations are badly resolved,
which means that most of the ionising radiation is absorbed within $5$
cell widths from the stellar sources, and hence the direct radiation
pressure is potentially underestimated. 

We have therefore used the aforementioned reduced flux approximation,
re-normalising ${\bf F}$ to $\cred E$ for the radiation pressure force
in each cell, i.e. assuming a full reduced flux, in the bulk direction
of the radiation. As discussed in \Sec{methods.sec}, this means we
\emph{overestimate} the radiation pressure in two types of locations:
i) In cells hosting stellar radiation sources, where the radiation is
in reality isotropic, but we instead take it all to point in the same
(average) direction.  ii) In-between radiation sources, where the
radiation pressure from opposing fields of radiation would in reality
cancel out, but again we instead take it to point in the average
direction, which is the direction away from the strongest
source. Since we found the effect of direct radiation pressure to be
negligible, the use of the reduced flux approximation is conservative
and our conclusions are robust.

\section{Resolution tests} \label{restests.sec}
\begin{figure}
  \centering
  \includegraphics[width=0.49\textwidth]
    {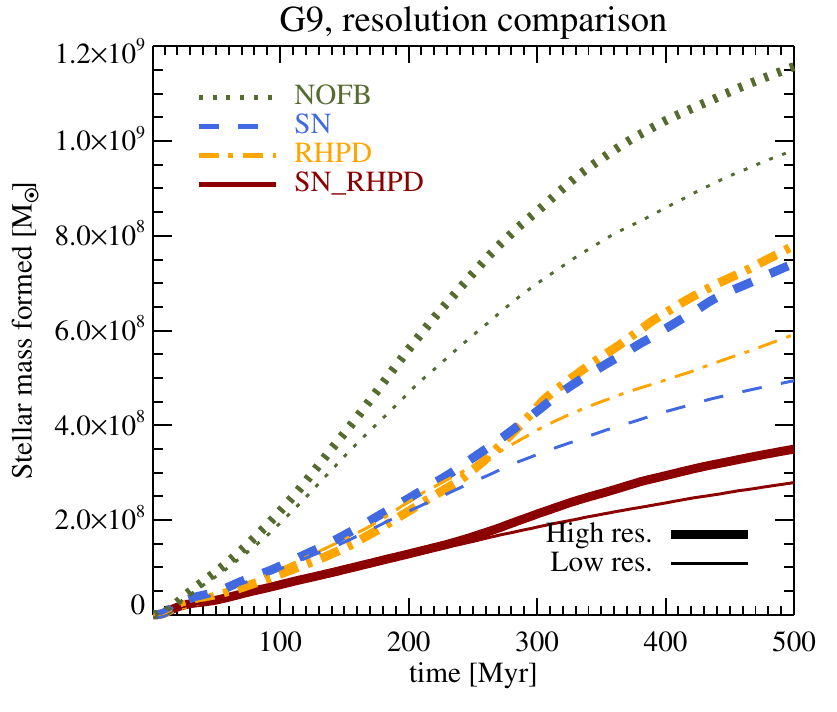}
  \caption
  {\label{sf_G9_lr.fig}Resolution tests. The plot shows comparisons
    of the stellar mass formed in the \sbc{} galaxy for default (thick
    curves) and low (thin curves) resolution, for the various feedback
    models, as indicated in the legend.}
\end{figure}

\begin{figure}
  \centering
  \includegraphics[width=0.49\textwidth]
    {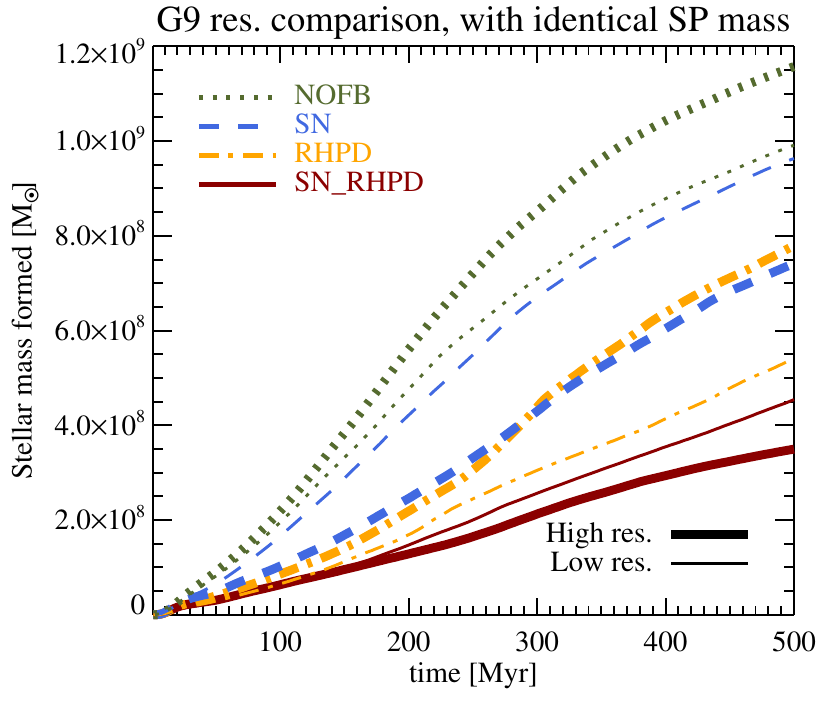}
  \caption
  {\label{sf_G9_lr_sm.fig}Resolution tests with constant stellar
    particle mass ($600 \ \Msun$). The plot shows comparisons of the
    stellar mass formed in the \sbc{} galaxy for default (thick
    curves) and low (thin curves) resolution, for the various feedback
    models, as indicated in the legend.}
\end{figure}

As is the case with simulation work in general, it is important to
investigate the dependence of the results on the numerical resolution.
For this reason, we compare runs of the fiducial \sbc{} galaxy with
lower-resolution counterparts, where the minimum cell width is two
times larger and the mass of particles (both formed and in the initial
conditions) is $8$ times larger\footnote{Correspondingly, the initial
  condition particles are $8$ times fewer.}. Otherwise the simulation
parameters are identical.

\Fig{sf_G9_lr.fig} shows the effect of resolution on star formation,
with line thickness indicating the resolution (thick for high
resolution and thin for low resolution), and as usual the colours and
linestyles represent different feedback models. We skip here the
results from the runs comparing the different radiation processes,
i.e. including/excluding direct and reprocessed radiation pressure,
and show only the all-inclusive radiation runs, but note that the
radiation pressure processes have little and seemingly random effects,
i.e. the dominant radiation effect is heating, as we have established
in the previous sections.

Lowering the resolution has the effect of reducing the formation of
stars, regardless of the feedback process included, even without any
feedback. This is a natural outcome of lowering the resolution, since
it becomes more difficult for the gas to collapse to high densities,
which in turn decreases the star formation rate which scales locally
as $\rho^{3/2}$. Indeed we find the mean densities typically to be
higher in the high-resolution runs than in their low-resolution
counterparts, by about half a dex. Apart from this systematic
suppression in star formation rates with resolution, stellar radiation
feedback reduces the formation of stars by roughly a similar fraction:
without SN feedback, the addition of radiation feedback reduces the
stellar mass formed at $500$ Myr by $\approx40\%$. Combined SN and RT
feedback results in very similar star formation for the two
resolutions, indicating numerical convergence.

In the resolution comparison, it is questionable whether $\mstar$
should be changed in the low resolution simulations. Increasing the
stellar particle mass by a factor of $8$, as we have done in
\Fig{sf_G9_lr.fig}, can boost the feedback, since each particle then
has eight times higher luminosity and SN energy, thus contaminating
the pure effect of changing the physical resolution. For the sake of
completeness, we thus also ran lower-resolution counterparts to the
\sbc{} simulations, exactly as just described, but with
$\mstar =600 \, \Msun$, identical to the fiducial simulations, for
which the star formation is compared to the higher-resolution case in
\Fig{sf_G9_lr_sm.fig}. Here the effect of SN feedback is negligible at
low resolution, while pure radiation feedback is \emph{more} efficient
than shown in \Fig{sfr_G9.fig}.

\section{Reduced light speed convergence tests} \label{ctest.app} 

To prevent a prohibitively small timestep in our RHD scheme, we use a
default reduced light speed of $\cred=c/200$ in our simulation
runs. This is in fact a six times lower light speed than recommended
in the analysis of reduced light speeds in ISM simulations in
\cite{Rosdahl:2013cea}, so it is important to verify that the chosen
light speed does not affect our results. For this purpose, we have run
the lower-resolution equivalent of the \sbc{} galaxy\footnote{eight
  times fewer/more massive particles and twice the minimum cell width
  compared to the default resolution, just as in \App{restests.sec}.}
with light speeds six times lower, two times higher, and six times
higher than the default value, the last value being recommended by
\cite{Rosdahl:2013cea}.

\begin{figure}
  \centering
  \includegraphics
    {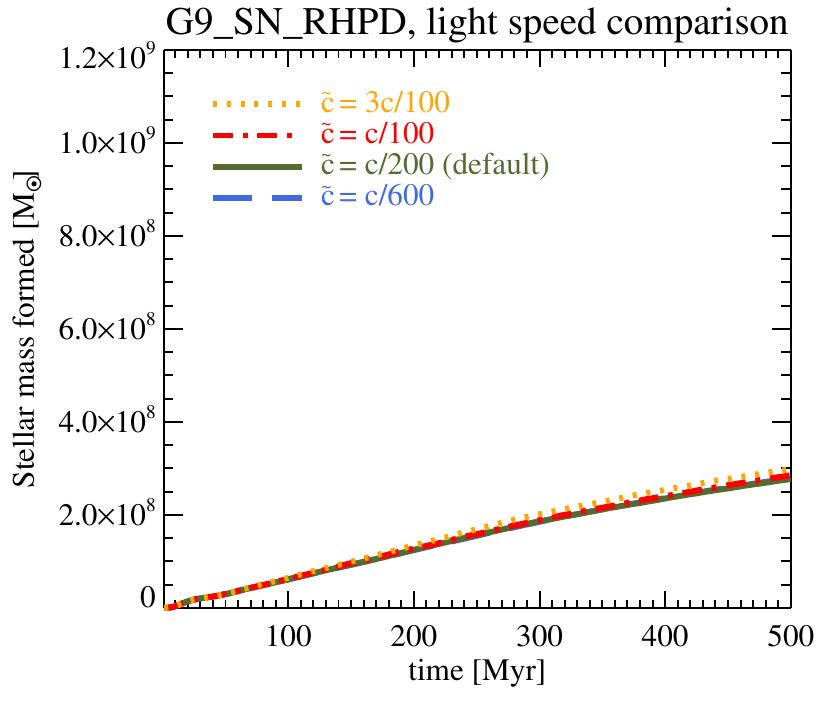}
  \caption
  {\label{sf_G9_cspeed.fig}Light speed convergence tests. The plot
    shows comparisons of the stellar mass formed in the \sbc{} lower
    resolution galaxy (see Sec. \ref{restests.sec}) for the default
    light speed of $c/200$ (solid green), along with identical runs
    with the light speed changed by factors of $6$, $2$, and $1/3$, as
    indicated in the legend.}
\end{figure}

In \Fig{sf_G9_cspeed.fig} we plot a comparison of the different light
speed runs in the form of the total stellar mass formed during the
$500$ Myr run time. The light speed has a negligible effect on the
star formation.  The morphology, outflow rates, and density
distributions are also nearly identical in the different light speed
runs.

We conclude that our results are well converged in terms of the
employed light speed, and we expect that similar results would be
retrieved with the full light speed\footnote{Apart from the problem of
  the computational cost of a run with the full light speed, such a
  simulation would also likely suffer from hydrodynamical diffusion
  with the current setup, due to a large number of very small
  timesteps. For a full light speed to work with our explicit RT
  solver, we would need to sub-cycle the RT within the HD step.}.

While true in the main simulation runs described in this paper, our
conclusion on light speed convergence does not necessarily hold when
the IR optical depth becomes very high, as in our ``extreme''
simulations described in \Sec{whatif.sec} where we artificially
boosted the IR opacity by orders of magnitude compared to the more
realistic theoretically motivated value, and found very reduced star
formation and outflows. When the optical depth becomes very high, the
effective propagation speed of radiation scales inversely with the
local optical depth, i.e. radiation waves travel at a speed $c/\tau$,
where $\tau$ is the optical depth across some relevant length scale,
such as an optically thick cloud (see e.g. sections 2.4 and 3.5 in
\RM{}). With our reduced speed of light, radiation waves travel at a
speed $\cred/\tau$, and if $\cred$ is orders of magnitude smaller than
the real light speed, as in this paper, the speed of light can become
a severe issue in very optically thick gas, with radiation waves
potentially travelling at a speed slower than the gas itself. Since
this becomes most severe with the highest optical depths, we ran the
most extreme experiment from \Sec{whatif.sec} ($\kOIR=10^4 \, \ccg$)
with $\cred$ increased by a factor two and decreased by a factor three
from the default value, i.e.  $\cred=10^{-2} \ c$ and
$\cred=1.67\times 10^{-3} \ c$, respectively. We found that the
average star formation rates are unaffected, but that they fluctuate
on a longer timescale with decreasing light speed, and that outflow
rates decrease very substantially with increasing light speed
(reinforcing our conclusion that radiation does not produce
outflows). It thus appears that with very large optical depth, the
reduced speed of light does become an issue for outflows, but not for
star formation rates. The main conclusions of this paper are not
affected though, since optical depths in our simulations are never
very high, save for the extreme ``what if'' scenario described in
\Sec{whatif.sec}.

\end{document}